# Advancements in Entangled Photon Pairs in 2D Van der Waals Materials for On-chip Quantum Applications

Abdus Salam Sarkar

Department of Physics, Stevens Institute of Technology, Hoboken, NJ, 07030, USA

Center for Quantum Science and Engineering, Stevens Institute of Technology, Hoboken, NJ, 07030, USA

Email: asarkar5@stevens.edu; salam.phys@gmail.com

**Abstract**

The next generation of technology is rooted in quantum-based advancements. The entangled photon pair sources play a pivotal role in such wide range of advancement in quantum-based applications, including quantum high precision sensors, communication, computing, cryptography and so on. Scalable on-chip quantum photonic devices hold the key potential to drive transformative advancements in the field. This review article highlights recent breakthroughs and advancements in the generation of entangled photon pairs (EPP) in two dimensional (2D) van der Waals (vdW) materials, with a focus on their applicability to quantum technologies and plausible on-chip quantum integrated photonic device technology. The article begins by discussing the fundamental principles of entangled photon pairs generation. It provides a comprehensive review of the origin and generation of entangled photons in emerging layered vdW materials, alongside various optical quantum characterization techniques. The review then explores key physical parameters of the quantum states associated with entangled photon pairs. Additionally, it examines concepts related to the realization of paired photon generation at the quantum limit. The final section focuses on the potential for on-chip integrated quantum device applications. Beyond highlighting recent advancements in quantum-based research, the review also outlines the current limitations and future prospects aimed at advancing the field.

## 1. Introduction

Quantum optics-based technology is rapidly advancing, ushering in a new era of technological innovation.[1–5] In particular, the emerging quantum applications such as secure quantum key distribution,[6–8] quantum communication,[9] quantum sensing,[10] quantum metrology,[8,11] quantum imaging,[12] and distributed quantum computing schemes[13] are actively testing and pushing the boundaries of fundamental quantum mechanics. Entangled photon-pairs (EPP), a fundamental concept in quantum physics, are critical sources and can enable the practical implementation of this advanced technology. A wide variety of entangled photon sources (EPS) have been extensively explored,[14–17] often relying on spontaneous parametric down conversion (SPDC) in the second harmonic nonlinear crystals. The second order nonlinear ($\chi^{(2)}$) process is the core workhouse for generating entangled quantum light states. The second order non-linear process, which typically occurs in a non-centrosymmetric material with ($\chi^{(2)} \neq 0$).[18–22] Conventionally, bulk crystals such as beta barium borate (BBO), potassium titanyl phosphate (KTP), and lithium niobate (LN) are used for SPDC.[1,2,21,23] Most recently, micro resonators and on-chip waveguides have been developed using periodically poled lithium niobate (PPLN) and LN microdisks, which show promise as bright SPDC sources. However, these materials typically hold significantly longer light-matter interaction lengths ($d$) than their coherence lengths ($\lambda_c$). As a result, satisfying the phase-matching (PM) condition is essential for achieving high efficiency and brightness in SPDC.[24] In particular, when $d<\lambda_c$, the generated entangled photon pairs (EPPs) are in phase, whereas at very small $d$, a large phase mismatch is tolerated. As a result, this leads to a broad spectrum of emitted photons.[24] Moreover, during the nonlinear process in a thick crystal, chromatic dispersion introduces differences in wavevectors, resulting in destructive interference among the parametric waves generated at different positions within the crystal, thereby reducing the conversion efficiency. Nonetheless, the bulk crystal imposes restrictions on the angle and frequency of the EPPs.

Two-dimensional (2D) atomically thin layered materials exhibit extraordinary light-matter interactions, including strong optical nonlinearity, due to confined electronic states and reduced dielectric screening (**Figure 1a**). Since the first isolation of graphene, 2D van der Waals (vdW) materials have been extensively studied for their nonlinear optical properties.[19,25] These atomically thin layered materials exhibit remarkable characteristics, particularly in terms of light-matter interactions length ($d \ll \lambda_c$) and enhanced nonlinear responses (**Figure 1c**). The crystal symmetry of these materials plays a pivotal role in determining their nonlinear behavior and corresponding coefficients. For instance, graphene and transition metal dichalcogenides (TMDs) typically possess 2H and 3R crystal structures, which are centrosymmetric (**Figure 1b**). In contrast, materials like black phosphorene and certain metal chalcogenides exhibit lower crystal symmetries, such as $D_{2h}$ and $C_{2v}$, which are non-centrosymmetric. These symmetry features (**Figure 1d**) make them highly suitable for nonlinear, on-chip quantum optical applications, offering a promising high-performance platform for subwavelength SPDC sources. In contrast, vdW materials with the *C*2 space group are gaining significant attention due to their exceptional second harmonic generation (SHG), making them excellent sources for SPDC photons.[18,26–31] The broad spectral response of vdW materials (**Figure 1a**) even at monolayer thickness, makes them ideal candidates for exploring hyperentanglement in the frequency domain. Moreover, these materials are inherently compatible with nanoscale devices and are well suited for integration into nanophotonic on-chip industrial applications.[32–36] Various 2D vdW materials have exhibited significantly enhanced nonlinear optical responses compared to conventional bulk materials, and demonstrating strong potential as ultrathin sources for SPDC. For example, Guo et al.[28]

introduced vdW niobium oxychloride ($NbOCl_2$) as a promising candidate for SPDC applications. The favorable susceptibility tensors within the crystal structure play a pivotal role in enabling SPDC and the generation of entangled quantum states. In another study, polarization entangled quantum states were successfully realized using an engineered vdW bilayer with an orthogonal stacking configuration an architecture not accessible in $NbOCl_2$ crystals.[27] Additionally, stacked vdW crystals have been shown capable of generating two or more complex quantum states at the nanoscale. A particular notable achievement is the realization of micrometer scale entangled photon pair sources in 3R-stacked TMD crystals,[23,37] attributed to the structural integrity of the 3R phase. Beyond this, twist-angle engineering in 2D materials can introduce a tunable nonlinear geometric phase, which can be utilized to compensate for phase mismatch in the SHG process.[38]

The second order nonlinear susceptibility $\chi^{(2)}$ of 3R-phase $MoS_2$ has been measured to exceed 800 pm/V,[39] which is comparable to or significantly higher than that of widely used nonlinear crystals such as BBO (3.9 pm/V) and KTP (29.2 pm/V).[40,41] This excellent nonlinearity effect in multilayer 2D vdW stacks are periodically modulated, offering a new degree of control for nonlinear optical engineering.[42] Nonetheless, 2D transition metal dichalcogenides (TMDs) and rhombohedral boron nitride (r-BN) has emerged as an excellent source of nonlinear optical signals.[17,38,43] r-BN exhibits broken inversion symmetry along both its armchair (AC) and zig-zag (ZZ) directions, enabling strong nonlinear optical activity. Notably, the nonlinear signal strength increases with the number of layers.[17] Generation of polarization EPPs has been successfully demonstrated in 2D r-BN, with a measured coincidence rate of 120 Hz and a coincidence-to-accidental ratio (CAR) of 200, corresponding to a fidelity of 0.93. A recent study reported an entangled photon pair generation rate as high as 3100 Hz/(mW×mm), achieved on a tunable platform for Bell state generation by varying the pump polarization without compromising entanglement quality. The highest reported fidelity of the polarization entangled state reached an exceptional level, demonstrating the quality and robustness of the generated entanglement. **Figure 1e** presents the timeline of the discovery of 2D van der Waals materials and the generation of photon pairs.

In parallel, other quasiparticle phenomena beyond entangled photon pairs are actively being explored in 2D vdW materials. These include electron-hole pairs (e-h),[44], excitons,[45–47] magnetic excitons (m-e),[48] magnon-magnon interactions,[49–51] and magnon-photon-phonon coupling,[52] chiral phonons,[53,54] electron-phonon (e-p) interactions,[55] single-photon sources and chiral phonon dynamics,[14] as well as valley physics,[47,56–58] and anti-parity-time (a-p-t) symmetry phenomena,[59] are also gaining momentum. Despite this broad progress, the generation of entangled photon pairs in vdW materials stands out as a key advancement for next-generation, on-chip integrated quantum technologies. This is due to the inherently rich nonlinear quantum optical properties enabled by these layered vdW materials.

This review presents the latest developments in the generation of quantum-EPPs in emerging vdW materials. The underlying physical mechanisms of entanglement via SPDC and the associated nonlinear optical properties are highlighted. Special attention is given to the influence of crystallographic structure on nonlinear quantum processes and its critical role in enabling SPDC. The realization of entangled quantum states in vdW materials is also discussed, including quantum state tomography, as well as the evaluation of state quality and purity for integrated on-chip quantum device applications. In addition to demonstrating the potential and significance of EPP generation in vdW systems, this article outlines current limitations and explores emerging opportunities and future prospects in the field.

## 2. Nonlinear optics crystallography and entanglement photon pairs generation

Current methods for generating EPPs include parametric amplification,[31] four-wave mixing,[60] and SPDC across various platforms.[61] The nonlinear crystals and quantum dots have emerged as key materials at the forefront of this research. In these material systems, nonlinear optical effects play a pivotal role in the generation of entangled photon pairs. In particular, the crystallographic structure of a material is fundamental to the emergence of nonlinear effects. Nonlinear optical effects arise when the material's polarization (P) no longer responds linearly to an applied electric field (E). Specifically, the polarization is described by the following expansion:

$$P(t) = \epsilon_0\chi^{(1)}E(t) + \epsilon_0\chi^{(2)}E^2(t) + \epsilon_0\chi^{(3)}E^3(t) + \cdots \ldots + \epsilon_0\chi^{(n)}E^n(t) \quad (1)$$

where, $\epsilon_0$ denotes the vacuum permittivity, $\chi^{(1)}$ is the linear (first order) susceptibility, and $\chi^{(n)}$ represents the $n^{th}$-order nonlinear susceptibility of the material for n>1. **Equation 1** can expressed in the frequency domain, where $\chi^{(n)}(\omega)$ denotes the $n^{th}$ order nonlinear susceptibility at frequency ω.

Nonlinear optical processes are generally classified as either parametric or nonparametric. In a parametric process, energy exchange occurs exclusively among interacting photons, and the total photon energy is conserved. In this case, the real part of the susceptibility $\chi$ is relevant. Additionally, both linear and angular momentum of the optical fields remain unchanged between the initial and final states. Since the quantum state of the material remains unaltered, the process occurs on an extremely short time scale governed by the energy-time uncertainty principle, ($\Delta t$=ђ/2$\Delta E$). Such processes are responsible for frequency conversion phenomena. A prime example is SHG, where frequency doubling occurs through a virtual intermediate state. On the other hand, nonparametric processes involve energy exchange between photons and the material itself. In these processes, the imaginary component of the susceptibility χ becomes significant (i.e., non-zero). The physical mechanism entails real energy transfer between the *initial* and *final* states of the system. Because this energy exchange involves transitions between real quantum states, nonparametric processes typically occur on longer timescales compared to parametric processes. Examples of such processes include two-photon absorption, saturable absorption (SA), and stimulated Raman scattering (SRS).

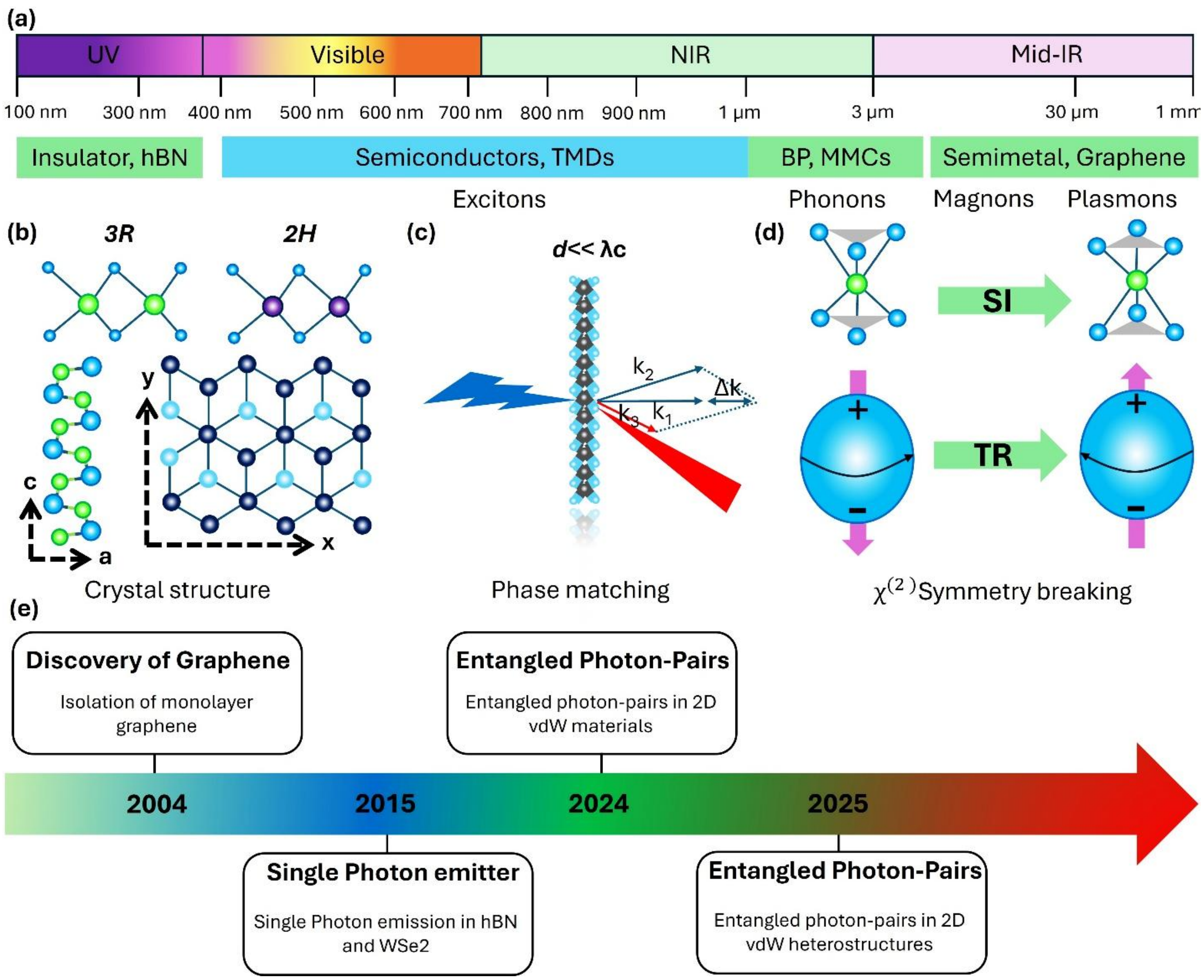

**Figure 1|** Key features of atomically thin layered materials for nonlinear optics and quantum entanglement. **a**, The electromagnetic spectrum and the unique optical properties of 2D materials across a broad spectral range. Distinct optical features arise from excitations such as excitons,

phonons, plasmon and magnons. **b**, Crystal structures of layered materials. *Top:* Different symmetries of an atomically thin 2D TMD of $MoS_2$, including hexagonal (2H) and rhombohedral (3R) phases. *Bottom left:* Crystal structure of an anisotropic atomically thin monochalcogenide (MX), where blue atoms represent M (metal) and green atoms represent X (chalcogens). The *a* and *c* directions denote the armchair and zigzag directions, respectively. *Bottom right:* Crystal structure of rhombohedral boron nitride (r-BN), with *x* and *y* representing the armchair and zigzag directions of the crystal. **c**, Schematic illustrating the impact of short propagation lengths ($d \ll \lambda_c$) on phase-matching conditions. $k_1$ and $k_2$ are wave vectors of two photons and $\Delta k = k_2 - 2k_1$. **d**, Symmetry breaking in atomically thin crystals, involving both spatial (SI) inversion and time-reversal (TR) symmetry and **e** A historical timeline of 2D van der Waals materials and the development of entangled photon pair generation.[28,37,62–64]

Here, primary focus is on second-order nonlinearity and the influence of crystallographic effects on this process. In **Equation 1**, the polarization P includes a term proportional to $E^2$, indicating that second-order nonlinear effects arise when the induced polarization is proportional to the square of the applied electric field ($P \propto E^2$). The second-order susceptibility, $\chi^{(2)}$, plays a major role in these nonlinear optical processes. In particular, $\chi^{(2)}$ is essential for SHG in crystalline materials. SHG is only permitted in materials that lack inversion symmetry i.e., those with a non-centrosymmetric crystal structure. In centrosymmetric layered materials, $\chi^{(2)}$ vanishes, making second-order nonlinear processes such as SHG forbidden. Since the 2$^{nd}$ order nonlinear processes require broken inversion symmetry in the crystal, understanding symmetry is essential. **Figure 1d** illustrates the symmetry breaking present in various crystals, including both spatial inversion (SI) symmetry and time-reversal (TR) symmetry. A wide range of materials has been studied to explore how symmetry influences nonlinear optical processes such as SHG.[18,19,65] For example, TMDs exhibit layer number dependent SHG ($I_{SHG} \propto N^2$).[22,66] In the 2H phase, odd numbered layers break inversion symmetry and therefore support SHG. In contrast, even numbered layers restore inversion symmetry, resulting in the disappearance of SHG. In the case of the 3R phase (rhombohedral stacking), the crystal structure inherently lacks inversion symmetry, making it non-centrosymmetric. As a result, 3R-phase TMDs exhibit non-zero SHG, regardless of the number of layers.

In theoretical background,[38,67] 2D materials possess in-plane three-fold rotational symmetry (RS) and mirror symmetry (MS). The nonzero component of $\chi^{(2)}$ tensor is $\chi_{xxx} = -\chi_{xyy} = -\chi_{yyx} = -\chi_{yxy} = \chi$. The *x*- and *y*- directions correspond to the AC and ZZ orientations of the crystal, respectively. The corresponding polarization equations are expressed as $P_x = \epsilon_0\chi\left(E_x^2 - E_y^2\right)$ and $P_y = -2\epsilon_0\chi E_x E_y$. However, in the case of circular polarizations, $P_L = \sqrt{2}\epsilon_0\chi E_R^2$ and $P_R = \sqrt{2}\epsilon_0\chi E_L^2$ where P= $(P_L, P_R)^T$, and *L* and *R* refer to the left- and right- circular polarization (LCP and RCP) states of light, respectively. Under a rotation operation defined as R=$\begin{pmatrix} e^{i\theta} & 0 \\ 0 & e^{-i\theta} \end{pmatrix}$ the polarization evolves as $P_L' = \sqrt{2}\epsilon_0\chi E_R^2 e^{i3\theta}$ and $P_R' = \sqrt{2}\epsilon_0\chi E_L^2 e^{-i3\theta}$. This rotation operation introduces a phase shift of 3θ in the SHG polarization. However, it does not affect the amplitude of SHG polarization. This phase shift is directly linked to the crystal symmetry and is known as the nonlinear Pancharatnam-Berry phase.[68,69] This phase serves as a key parameter for compensating phase mismatch during the phase-matching process in nonlinear crystals.

Materials with a puckered or wavy lattice structure exhibit reduced crystal symmetry such as the ($D_{2h}$) point group compared to conventional 2D materials like graphene ($D_{6h}$) and TMD.

This reduction in symmetry leads to more intriguing nonlinear optical phenomena. The lower symmetry of these crystals results in broken inversion symmetry and introduces in-plane anisotropy. Additionally, the buckled structure gives rise to direction dependent optical properties, making these materials particularly promising for polarization sensitive and anisotropic nonlinear optical applications. Most recent development of monolayer 2D monochalcogenides (MMCs) have been shown to exhibit low crystal symmetry belonging to the $C_{2\upsilon}$ point group, with an orthorhombic crystal structure (Pnma space group).[70–74] In this structure, inversion symmetry is further broken, enhancing their nonlinear optical response. This unique crystallographic feature gives rise to novel optical polarization properties, enabling the observation of new orders of optical nonlinearity and anisotropy.[73,75–78]

On another note, 2D magnetic materials have also exhibited SHG,[20,79–85] which classified into two types: *c*-type and *i*-type.[86,87] Among these, *c*-type materials have been more widely reported to show SHG, primarily due to their non-centrosymmetric antiferromagnetic (AFM) nature. Interestingly, even centrosymmetric AFM materials have demonstrated giant SHG responses, attributed to their magnetic structures, which break both SI and TR symmetries.[80]

## 2.1 Quantifying Nonlinear Coefficients and Conversion Efficiency

The nonlinear coefficient and conversion efficiency are key metrics for evaluating the performance of nonlinear optical materials.[41] In atomically thin layered materials, values of $\chi^{(2)}$ have been reported in the order of nanometers per volt (nm/V), which is significantly higher than those observed in conventional nonlinear crystals, where $\chi^{(2)}$ is typically on the order of picometers per volt (pm/V). In particular, Group IV monochalcogenides have demonstrated remarkably high $\chi^{(2)}$ values. However, these values vary widely among layered materials due to several influencing factors, including material quality, fabrication methods, strain, stress, and doping levels. The intensity of SHG, $I_{2\omega}$, is estimated by

$$I_{2\omega} \ \alpha \ I_0^2 l^2 sinc^2\left(\frac{\Delta k.l}{2}\right) \tag{2}$$

where $I_0$ is the input intensity, $l$ is the propagation length, $c$ is the speed of light and $\Delta k$ is the phase mismatch. For an angled beam or co-linear propagation inside the crystal, the phase mismatch for perpendicular (normal) incidence is given by $\Delta k = k_{2\omega}cos\theta_{2\omega} - 2k_{\omega}cos\theta_{\omega}$. For in-plane (parallel) incidence, it is given by $\Delta k = k_{2\omega} sin\theta_{2\omega} - k_{\omega} sin\theta_{\omega} = k_{2\omega} sin\theta_{2\omega} - 2k_{\omega} sin\theta_{\omega}$, where $\theta$ is the propagation angle. Ideally, efficient frequency conversion requires $\Delta k$=0. This signifies that achieving higher SHG efficiency in 3R-$MoS_2$ for $l >> \lambda_c$ requires phase matching. For example, the estimated $I_{2\omega}$ of monolayer $MoS_2$ at 800 nm is approximately $10^{-7}$ (**Figure 2b**), while $WSe_2$ exhibits a lower value on the order of $10^{-10}$. For instance, the angle of incidence plays a crucial role in on-chip integrated quantum device applications. In particular, both in-plane and out-of-plane incidence significantly affect SHG efficiency and the corresponding phase-matching conditions. In 2D nonlinear materials, in-plane SHG is typically generated under normal incidence in the far field, whereas out-of-plane SHG can be enabled through edge coupling in a waveguide geometry.[99] However, it is well established that efficient in-plane momentum propagation can be achieved by using waveguide-integrated prism or grating couplers placed directly on top, which is beneficial for enhancing coupling efficiency. Notably, SnS and SnSe have demonstrated extraordinarily high SHG intensities, with the highest $I_{2\omega}$ values estimated to be for SnS and SnSe, respectively. The key parameters, including $\chi^{(2)}$, for various emerging layered materials are summarized in **Table 1**.

**Table 1.** *The 2nd order nonlinear coefficient for the vdW materials. ME is mechanical exfoliation, LPE is liquid phase exfoliation, CVT is chemical vapor transport, PVD is physical vapor deposition, VLSS is vapor-liquid-solid synthesis method and MSAA is metal solvent assisted approaches.*

| Materials | Prep Method | Substrate | Layer number | SHG Wavelength (nm) | $\chi^{(2)}$ | Reference |
|---|---|---|---|---|---|---|
| **$MoS_2$** | ME | $SiO_2$ | Single layer | 405 | 10000 pm/V | [88] |
| | CVD | $SiO_2$ | Single layer | 405 | 5000 pm/V | [88] |
| | ME | $SiO_2$ | Single layer | 950 | 800 pm/V | [39] |
| **$MoSe_2$** | ME | Si waveguide | Single layer | 775 | 7800 pm/V | [89] |
| | CVD | $SiO_2$ | Single layer | 810 | 50 pm/V | [90] |
| **$WS_2$** | ME | $SiO_2$ | Single layer | 416 | 4500 pm/V | [91] |
| **$WSe_2$** | ME | $SiO_2$ | Single layer | 408 | 5000 pm/V | [92] |
| **h-BN** | ME | Fused Silica | Multilayer (36.2 nm) | 406 | 42 pm/V | [93] |
| | ME | Quartz | Few layer | 405 | $200\times10^{-11}$ m/V | [65] |
| | CVD | Sapphire | Single layer | 1046 | 83.2 pm/V | [94] |
| **r-BN** | *VLSS* | fused silica | 730 nm | 800 | 30 pm/V | [43] |
| **SnS** | PVD | Mica | Multilayers (7 nm) | 800 | 100 pm/V | [78] |
| **$NbSe_2$** | ME | Quartz | 1L<br>3L | 1030 | $1.0\times10^3$ pm/V<br>~73 pm/V | [95] |
| **$NbOCl_2$** | CVD | - | Single Crystal | 532 | 160 pm/V | [27] |
| **$NbOBr_2$** | CVT | Si | 10 nm | 1500 | $9.16\times10^{-11}$ m/V | [18] |
| **$NbOI_2$** | CVT | Quartz | multilayer | 425-525 | $10^{-9}$ m/V | [96] |
| **Perovskite** | Chemical | - | - | 850 | 0.68 pm/V | [97] |
| **MOF** | Chemical synthesis | Quartz | 12.2 μm | 1030 | ~19.86 pm/V | [98] |

### 2.2 Spontaneous parametric down conversion

Spontaneous parametric down-conversion (SPDC) is a nonlinear optical process, also known as a form of difference frequency generation (DFG).[61,100] In other words, SPDC is the inverse of SHG and adheres to the same energy-sum rule. While SHG is a coherent frequency upconversion process, SPDC acts as its reverse, producing quantum-correlated or entangled photon pairs. In this process, a high energy pump photon spontaneously decays into two lower energy photons, referred to as the *signal* ($s$) and *idler* ($i$) photons (**Figure 2a**). This process conserves both energy ($\omega_1 = \omega_2 + \omega_3$; 1 and 2 correspond to *signal* and *idler*) and momentum ($k_1 = k_2 + k_3$; 1 and 2 correspond to *signal* and *idler*) (**Figure 2a**). The *signal* and *idler* photons are always produced as a pair, and they can become entangled in various degrees of freedom such as polarization, energy-time, or position-time. **Figure 2a** illustrates the different types of SPDC. Type-I SPDC includes both degenerate and non-degenerate processes. In degenerate Type-I SPDC, the *s/1* and *i/2* photons are identical and exit the crystal at equal but opposite angles. In contrast, in the non-degenerate case, the exit angles depend on the wavelengths of the *s* and *i* photons. This results in a set of colored rings centered around the pump beam (in a cross-sectional view). On the other hand, **Figure 2a** (bottom right) shows that Type-II SPDC produces two distinct rings corresponding to the signal and idler photons. When the *signal* and *idler* photons have orthogonal polarizations, the system can form a polarization entangled Bell state, which is characteristic of Type-II SPDC (**Figure 2a bottom**). Therefore, SPDC serves as a fundamental mechanism for generating entangled photon pairs and plays a vital role in many quantum optics and quantum information experiments. Moreover, the ultrathin vdW materials offer great potential as optically enabled sources of entangled photons through SPDC.

To understand the dynamics of the generated photons as *signal* and *idler*, the Green's Function Quantization (GFQ) method is employed.[101] Using the GFQ framework, SPDC is analyzed in a nonlinear (NL) material excited by a single-frequency pump $E_p(r,t) = E_p(r)e^{\{-i\omega_p t\}} + c.c.$ In the case of far-field coincidence detection with spatial three-dimensional (3D) coordinates, the function described by

$$\mathcal{R}(\boldsymbol{r}_s, \boldsymbol{r}_i,\ \boldsymbol{e}_s, \boldsymbol{e}_i, \omega_s, \omega_p - \omega_s) = \left(\frac{8}{\pi} n_i n_s r_i^2 r_s^2 \frac{(\omega_p-\omega_s)^3\omega_s^3}{c^6}\right) \times \left|\sum_{\alpha,\beta,\gamma} \sum_{\sigma_s,\sigma_s} e_{i,\sigma_i} \int_V \ d\boldsymbol{r}\, \chi^{(2)}_{\alpha\beta\gamma}(\boldsymbol{r}) \times E_{p,\gamma}(\boldsymbol{r}) G_{\sigma_s\alpha}(\boldsymbol{r}_s, \boldsymbol{r}, \omega_s) G_{\alpha_i\beta}(\boldsymbol{r}_i, \boldsymbol{r},\ \omega_p - \omega_s\right|^2 \quad (3)$$

The detection rate is defined as $\mathcal{R} \equiv \frac{\mathrm{d}^4 N}{\mathrm{d}t\mathrm{d}\Omega_s\mathrm{d}\Omega_i\mathrm{d}\omega_s}$, which represents the number of photon pairs collected per unit time, per unit solid angle for the signal ($\mathrm{d}\Omega_s$) and idler ($\mathrm{d}\Omega_i$), and per unit angular frequency of the signal photon ($\mathrm{d}\omega_s$). The angles $\alpha, \beta, \gamma$, $\sigma_s$, and $\sigma_i$ span the spatial coordinates of *x*, *y* and *z* directions, respectively. Here, $G_{ij}(\mathbf{r}, \mathbf{r}', \omega)$ is the tensor component of the electric GF, and $\chi^{(2)}_{\alpha\beta\gamma}(\boldsymbol{r})$ is the 2nd order nonlinear susceptibility tensor. $\boldsymbol{r}_s$ and $\boldsymbol{r}_i$ denote the detection positions of the *signal* and *idler* photons, respectively, while $n_s$ and $n_i$ are the corresponding refractive indices at those positions.

By exploiting the homogeneity in the *x* and *y* directions in the spatial frequency domain, the GF can be expressed as an inverse Fourier transform[102]

$$\overleftrightarrow{\mathbf{G}}(\mathbf{r}, \mathbf{r}', \omega) = \frac{1}{(2\pi)^2} \int_{\infty}^{\infty} \mathrm{d}\mathbf{q}\, \overleftrightarrow{\mathbf{g}}(\mathbf{q}, \mathrm{z}, \mathrm{z}', \omega) \mathrm{e}^{\mathrm{i}\mathbf{q}\cdot(\mathbf{r}_\perp - \mathbf{r}'_\perp)} \quad (4)$$

where, $\mathbf{q}$ and $\mathbf{r}_\perp$ represent the 2D real wavevector and spatial vector in the transverse directions, respectively.

For a classical pump field under the undepleted pump approximation, the electric field is given by

$$\mathbf{E}_\mathrm{p}(\mathbf{r}) = \frac{1}{(2\pi)^2} \int_{\infty}^{\infty} \mathrm{d}\mathbf{q}\, \mathbf{E}_p(\mathbf{q}, \mathrm{z}) \mathrm{e}^{\mathrm{i}\mathbf{q}\cdot\mathbf{r}_\perp} \quad (5)$$

By introducing (B) and (C) in eq (A), one can separate the spatial integral into $\mathrm{d}\mathbf{r} \rightarrow \mathrm{dzd}\mathbf{r}_\perp$. The solution becomes, $\int_{\infty}^{\infty} \mathrm{d}\mathbf{r}_\perp\, \mathrm{e}^{\mathrm{i}(\mathbf{q}_p - \mathbf{q}_s - \mathbf{q}_i)\cdot\mathbf{r}_\perp} = (2\pi)^2 \delta(\mathbf{q}_p - \mathbf{q}_s - \mathbf{q}_\mathrm{i})$ (6)

The joint detection rate (JDR) or the count rate of detecting a pair of photons simultaneously within a given coincidence time window

$$\mathcal{R}\left(\mathbf{r}_s, \mathbf{r}_i, \mathbf{e}_s, \mathbf{e}_i, \omega_s, \omega_p - \omega_s\right) = \left(\frac{8}{\pi} n_i n_s r_s^2 r_i^2 \frac{(\omega_p-\omega_s)^3 \omega_s^3}{c^6}\right) \times$$

$$\left|\frac{1}{(2\pi)^2} \int_{\infty}^{\infty} \int_{\infty}^{\infty} d\mathbf{q}_s d\mathbf{q}_i\, \tilde{\mathcal{R}}\left(\mathbf{q}_s, \mathbf{q}_i, z_s, z_i, \omega_s, \omega_p - \omega_s\right) \times e^{i(\mathbf{q}_s \cdot \mathbf{q}_{s\perp} + \mathbf{q}_i \cdot \mathbf{q}_{i\perp})}\right|^2 \quad (7)$$

Where, $\tilde{\mathcal{R}}\left(\mathbf{q}_s, \mathbf{q}_i, Z_s, Z_i, \omega_s, \omega_p - \omega_s\right) = \sum_{\alpha,\beta,\gamma} \sum_{\sigma_s,\sigma_s} e_{i,\sigma_i} \times \int_{-a}^{0} dz\, \chi_{\alpha\beta\gamma}^{(2)}(z)\, E_{p,\gamma}(\mathbf{q}_s + \mathbf{q}_i, z) \times$

$$g_{\sigma_s\alpha}(\mathbf{q}_s, z_s, z, \omega_s)\, g_{\alpha_i\beta}\left(\mathbf{q}_i, z_i, z, \omega_p - \omega_s\right) \quad (8)$$

**Equation (7)** measures the joint angular probability (JAP) amplitude of the entangled photons. In contrast, **Equation (7)** is used to calculate the detection rate of entangled photon pairs generated via the SPDC process. In the far-field regime, where detection occurs at distances much greater than the source dimensions (i.e. $\mathbf{r}_s \gg \mathbf{r}$ and $\mathbf{r}_i \gg \mathbf{r}$) the far-field approximation is applied. Under this condition, the JDR

$$\tilde{\mathcal{R}}\left(\theta_s, \theta_i, \varphi_s, \varphi_i, \mathbf{e}_s, \mathbf{e}_i, \omega_s, \omega_p - \omega_s\right) = \left(\frac{8}{\pi} \frac{1}{(2\pi)^4} n_i n_s \frac{(\omega_p-\omega_s)^3 \omega_s^3}{c^6}\right) \times \left|\tilde{\mathcal{R}}\left(\mathbf{q}_s, \mathbf{q}_i, \omega_s, \omega_p - \omega_s\right) k_{z,1s} k_{z,1i}\right|^2 \quad (9)$$

Here, $\mathbf{q}_s$ lies in the *x*-*y* plane for both the signal and idler photons, which describes the JDR as a function of the detection angles $(\theta_s, \theta_i, \varphi_s, \varphi_i)$ in spherical coordinates. In particular, the JDR is proportional to the squared absolute value of $\tilde{\mathcal{R}}$ multiplied of entangled photons *z* components of the wave vectors.

### 3. Entanglement in 2D vdW materials

Entangled photon pairs are a fundamental concept in quantum physics, particularly in quantum optics and quantum information science. When two photons are entangled, they exhibit correlations in their physical properties such as polarization, momentum, energy, parity, spin that

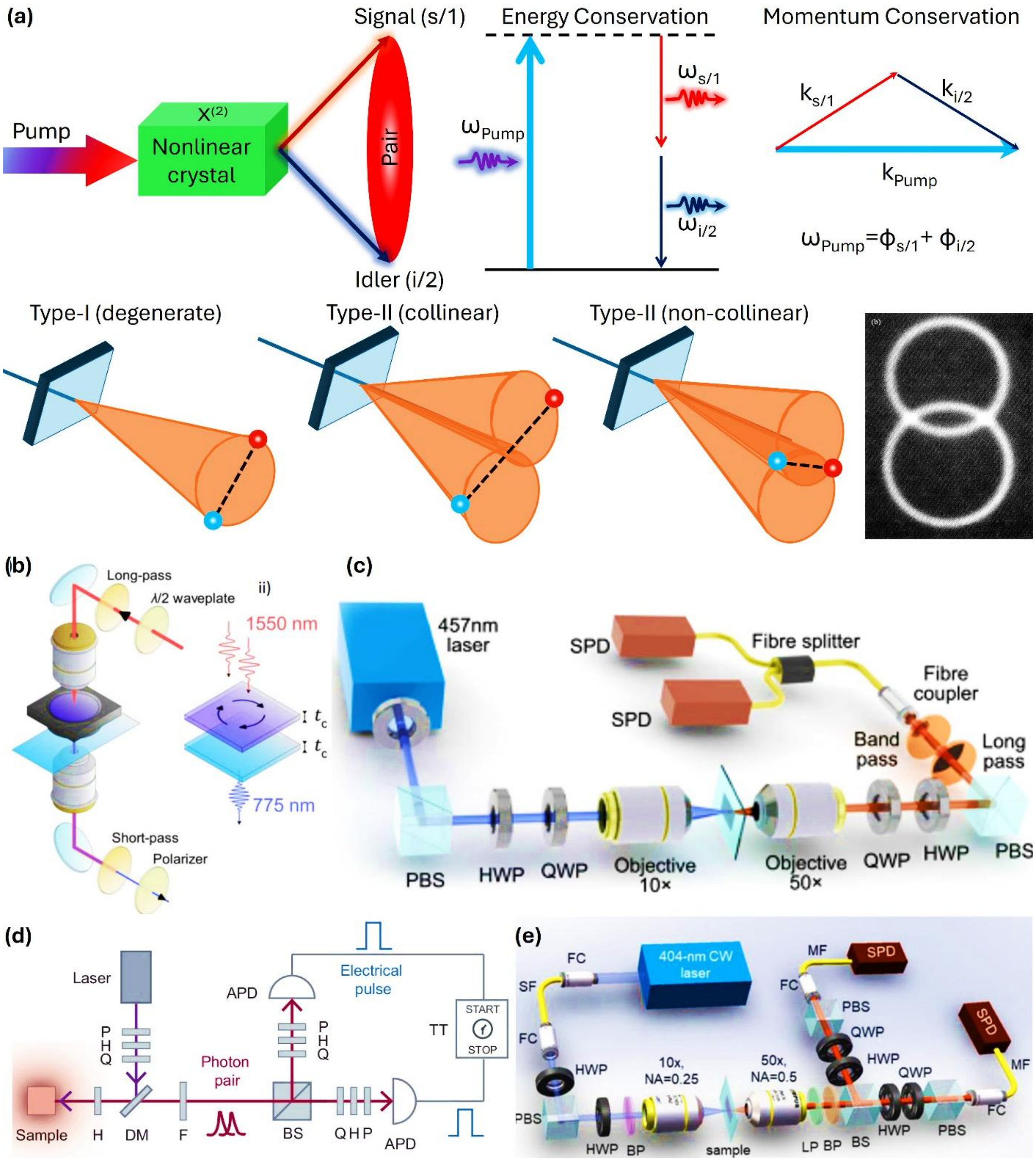

**Figure 2| Fundamentals and experimental setup for generating quantum-entangled photon**

**pairs and quantum state tomography. a**, Schematic illustration of the spontaneous parametric down-conversion (SPDC) process. A pump beam from a laser source interacts with a nonlinear crystal, producing pairs of photons known as the signal (s or 1) and idler (i or 2). The process adheres to the conservation laws of energy and momentum within the nonlinear medium. Energy conservation is expressed as $\omega_{Pump}=\phi_{s\ or\ 1} + \phi_{i\ or\ 2}$ where 1 and 2 denote the signal and idler photons, respectively. *k* represents the momentum vector components, which are also conserved during the interaction. *Bottom:* Schematic illustration of the fundamental SPDC process. Shown from left to right are degenerate Type-I SPDC, collinear Type-II SPDC, and noncollinear Type-II SPDC. *Bottom right:* Photograph of down-converted photons filtered at 702 nm with a 5 nm FWHM bandpass filter. **b,** Schematic of the quasi-phase-matching measurement setup. The measurements are performed with a 3R-$MoS_2$ crystal. Adopted from Springer Nature.[103] Copyright, 2024, CC BY-NC-ND 4.0. **c,** Schematic representation of the experimental optical setup for SPDC measurements. The entangled photon pairs are collected in transmission mode. Adopted from Springer Nature.[37] Copyright, 2024, Creative Commons Attribution 4.0 International License. **d,** Optical setup used for SPDC characterization by Lyu et al. A continuous wave (CW) laser is employed as the pump source. The setup includes polarizers, a dichroic mirror (DM), and a beam splitter (BS). The generated signal and idler photons are coupled into optical fibers, detected by avalanche photodiodes (APDs), and analyzed using a time tagger (TT). Reproduced with permission.[104] Copyright Creative Commons CC-BY-NC-ND 2025, Nature Springer. and **e,** Schematic illustration of the optical setup used for polarization quantum state tomography. Adopted from Gao et al.[23] Copyright 2024, Nature Springer. Creative Commons CC-BY-NC-ND.

are stronger than what is possible under classical physics, even when separated by large distances. This phenomenon is a direct consequence of quantum entanglement, a nonclassical link between particles that was famously discussed in the Einstein-Podolsky-Rosen (EPR) paradox.

Entangled photon pairs in vdW materials represent a frontier in quantum photonics, where 2D layered crystals such as TMDs, niobium oxide dihalides $NbOX_2$ (X = Cl, Br, I) enable compact and tunable quantum light sources. These materials exhibit strong nonlinear optical effects, excitonic effects and spin valley coupling, which can be harnessed to generate polarization entangled photon pairs (P-EPP) through processes such as biexciton exciton cascades. The atomically thin nature and high optical quality of vdW materials and heterostructures allow for precise control over light matter interactions, phase matching and their integration into photonic platforms makes them promise for on-chip quantum technologies.[103] Additionally, the tunability of band alignment and excitonic properties via stacking angle, strain, or dielectric environment enhances the potential as scalable sources for quantum communication and quantum information processing. The first demonstration of polarization-dependent entangled photon-pair generation in ultrathin layered van der Waals 3R-$WS_2$ was reported by Feng et al.[37] In particular, high photon-pair purity was achieved, with a high coincidence-to-accidental ratio (CAR).

### 3.1. Experimental setup to generate Entangled photon pairs

Experimental technologies serve a critical role in characterizing the quantum optical properties of the quantum systems. The following sections provide a detailed description of the measurement principles, the optical configurations, and the data acquisition methods. EPP generation is typically realized using the SPDC techniques. The experimental setup is designed to produce customized entangled photon pairs. In the basic configuration, a laser source is used to excite nonlinear material. The resulting photoexcited signal is filtered and then split into two paths. Each of these split beams passes through a combination of optical components. The signals are then collected at two ends using the photodetectors. To analyze the SPDC signal, coincidence measurements are carried out using a time tagger or time correlated single photon counting (TCSPC) system.

The first SPDC measurement in a layered van der Waals (vdW) material was reported by Feng *et al.,*[37] where entangled photon pairs were generated in monolayer $WS_2$ using the setup shown in **Figure 2c**. In this experiment, the photon pairs were pumped with LCP and RCP light. The same setup was employed to measure the coincidence rate as a function of the quarter-wave plate rotation in the collection path. Specifically, a Hanbury Brown and Twiss (HBT) setup is used to measure the second-order correlation function $g^{(2)}(\tau)$, revealing the photon statistics of the source. The basic principle involves splitting the incident photons into two spatially separated detection channels using a 50:50 free-space or fiber-coupled beam splitter. The arrival times of the photons at each single-photon detector following the beam splitter are monitored by a time tagger (TT) or a time-correlated single-photon counting (TCSPC) module. By correlating these detection events as a function of the time delay $\tau$, the setup can distinguish between classical and nonclassical photon emission characteristics. The optical path typically comprises the source output, spatial filtering (e.g., via a single-mode fiber), the beam splitter, and two detector arms leading to the TCSPC or TT unit (**Figure 2c, d**).

Quantum state tomography (QST) reconstructs the full density matrix $\hat{\rho}$ of a photonic quantum state by performing a complete set of projective measurements in different polarization bases (**Figure 2e**). In principle, this is achieved by rotating quarter- and half- wave plates (QWP/HWP) along with polarizers to project onto horizontal/vertical (H/V) and left/right circular polarization (LCP/RCP) states. The measurement statistics for each projection are collected, and numerical maximum-likelihood estimation is then applied to reconstruct $\hat{\rho}$, enabling the calculation of fidelity, purity, and entanglement measures. The optical path incorporates polarization-control elements before detection in one or both arms of the correlated photon pairs. The subsequent sections provide a detailed account of experimental observations in various vdW materials.

To characterize both the spectral and temporal properties of the generated photons, a spectrometer or single-photon detector (SPD) including superconducting nanowire single-photon detectors (SNSPDs) is combined with a TT/TCSPC system. The basic principle involves spectrally dispersing the photons and detecting them in coincidence with a timing reference, thereby revealing the joint spectral intensity (JSI) or JDR and spectral correlations between the signal and idler photons. The optical path begins at the photon source, passes through spectral filtering or dispersion optics, and terminates at a wavelength-resolved SPD, whose detection events are time-correlated. This approach enables the evaluation of spectral purity, indistinguishability, and energy-conservation constraints in parametric processes. Here, both transmission and reflection mode setups will be explored. In particular, for low-dispersion materials, SHG can achieve higher conversion efficiency in the transmission geometry compared to the back-reflection configuration.

For instance, Liang et al.[17] utilized a continuous wave (CW) laser at 405 nm was used to excite an 2D r-BN sample. The emitted signal photons passed through a sequence of filters, mirrors, and polarizers, before being coupled into an optical fiber via a fiber beam splitter. The split photons were then detected by avalanche photo diode (APDs), and coincidence measurements were conducted using a TCSPC system. On the other hand, a free space optical setup can be employed to collect SPDC photon pairs and perform coincidence measurements. The photoexcited

SPDC photon pairs are filtered and transmitted through a series of polarization optics. Instead of using a fiber coupled beam splitter, the photon pairs can separate into two paths using a free space beam splitter. A similar optical configuration has been demonstrated by Kallioniemi et al.[105] In all these setups, the generated SPDC photon pairs are collected in reflection mode with normal mode of incidence.

Qiu and co-workers[23] have presented the SPDC optical setup from multiple perspectives (**Figure 2e**). In their configuration, SPDC photon pairs are generated and collected in transmission mode from a nonlinear sample. Specifically, the nonlinear material is placed on a transparent PDMS substrate to facilitate signal acquisition. The generated SPDC photons are split into two distinct paths. These split photons pass through quarter- and/or half wave plates and a polarizing beam splitter (PBS), before being coupled into multimode fibers. Finally, the signal and idler photons are detected by two single photon detectors. The coincidence rate of the entangled photon pairs is measured using a TCSPC module.

Determining the total photon pair detection efficiency, $\eta_{tot}$, is a key measure of the overall performance of the setup. Specifically, $\eta_{tot}$ can calculate as the product of individual contributing factors[64] $\eta_{tot}=T_{opt}^2 \times T_{coupl}^2 \times \eta_{BS} \times \eta_{detec}^2 \times \eta_{LP}^2$. Where, $T_{opt}$ is the single photon optical transmission, accounting for losses from filters, mirrors, lenses, etc.; $T_{coupl}$ is the fiber coupling efficiency; $\eta_{BS}$ represents the non-uniformity and probabilistic splitting of the (fiber or nonfiber) beamsplitter; $\eta_{detec}$ is the detection efficiency of the detectors at the degenerate SPDC wavelength, averaged over different polarizations; and $\eta_{LP}$ is the spectral detection factor associated with the long pass filter. The $\eta_{tot}$ for photon pairs generated in TMDs is approximately 0.6%.

### 3.2. Entanglement in 2D vdW materials & heterostructures

The very first generation of P-EPP in 2D materials is realized by Qiu and co-workers.[37] In particular, an ultrathin rhombohedral tungsten disulfide (3R-$WS_2$) exhibits a high P-EPPs generation. Followed by this work, a compact and highly tunable EPPs source is explored in 3R-$MoS_2$ by Weissflog et al.[64]

In the first report, a thick layer of 3R-$WS_2$ crystal (*R3m* space group) was used to satisfy three stringent requirements for a compact entangled photon-pair (EPP) source: broken crystal symmetry, well-defined quantum entanglement, and a large second-order susceptibility $\chi^{(2)}$. The crystal exhibits ABC stacking along the out-of-plane direction, which breaks inversion symmetry from the monolayer to the bulk form. Consequently, the $\chi^{(2)}$ value of the crystal scales up significantly in thick layers (**Figure 3a**). As 3R-$WS_2$ belongs to the $C_{3v}$ crystal symmetry group characterized by a threefold rotational axis and mirror planes (**Figure 3b**) the non-zero 2nd order susceptibility tensor elements are $\chi^{(2)}_{yyy} = -\chi^{(2)}_{yxx} = -\chi^{(2)}_{xxy} = -\chi^{(2)}_{xyx} = |\chi^{(2)}|$. In the SPDC process with LCP/RCP pump light, the generated photon pairs are $|LL\rangle_{s,i}$ and $|RR\rangle_{s,i}$ with pump of $|R\rangle_{Pump}$ and $|L\rangle_{Pump}$, respectively. Through coherent superposition, the P-EPP state is $\frac{1}{\sqrt{2}}(|LL\rangle_{s,i} + e^{i2\alpha}|RR\rangle_{s,i})$ pumped with a linearly polarized state $\frac{1}{\sqrt{2}}(|R\rangle_{Pump} + e^{i2\alpha}|L\rangle_{Pump})$ at an incident angle $\alpha$. The P-EPP generation rate in SPDC was monitored via its power-dependent coincidence counts. For 3R-$WS_2$ with $C_{3v}$ symmetry, the coincidence rate displayed a linear dependence on pump power (**Figure 3c**). The SPDC selection rules for this crystal are defined by $|L\rangle_{Pump} \rightarrow |RR\rangle_{s,i}$ and $|R\rangle_{Pump} \rightarrow |LL\rangle_{s,i}$.

Furthermore, 3R-$MoS_2$ with a $C_{2v}$ phase crystal structure (**Figure 3d**) has been demonstrated as a core component of a compact and highly tunable EPP source.[64] This was evidenced by the generation of maximally entangled photon pairs from a microscale crystal. Specifically, 3R-$MoS_2$ enables broadband generation of maximally polarization-entangled Bell states at 1550 nm (**Figure 3e**), resolving EPPs in the telecom band. In thin, non-phase-matched crystals, the SPDC spectrum broadens, a feature observed in thick-layer 3R-$MoS_2$ crystals. Classically, polarization-resolved second-harmonic generation (P-SHG) reveals a six-fold symmetry pattern along the crystal axis (**Figure 3f**). This six-fold symmetry in the coincidence rate remains independent of pump polarization when SPDC detection is performed without a linear polarizer (**Figure 3g, h**). However, the pattern changes when the pump polarizer and analyzer are rotated. Notably, both parallel and perpendicular configurations exhibit a six-fold symmetry with a co-aligned pattern, following the form $R_{SPDC} \propto Sin^2(3\varphi_p)$, where $\varphi_p$ is the angle between the pump polarizer and the SPDC analyzer. By rotating the linear pump polarizer while keeping the analyzer at several fixed angles, the SPDC signals can be systematically verified. In the case of 3R-$MoS_2$, which possesses three-fold rotational crystal symmetry, a two-lobe pattern emerges (**Figure 3i-k**). For example, with the analyzer fixed at an angle of 110°, the SPDC reveals two lobes in the coincidence rate, which aligns well with the theoretically predicted polarization dependence with $R_{SPDC} \propto Sin^2(3\varphi_{pol} + \varphi_p)$.

Consequently, the rhombohedral phase boron nitride (r-BN) has garnered significant attention due to its unique symmetrical properties with superior brightness and CAR values (**Figure 3l**). Unlike conventional hexagonal boron nitride (h-BN), which exhibits a centrosymmetric crystal structure in even numbered layers thereby suppressing $2^{nd}$ order nonlinear optical processes r-BN intrinsically breaks inversion symmetry owing to its ABC-type interlayer stacking sequence. This broken symmetry is critical for enabling efficient second-order nonlinear processes, such as SHG, which are essential for quantum photonic applications including EPPs production. The diminished nonlinear response in centrosymmetric h-BN structures leads to a significantly reduced entanglement signal. In contrast, r-BN's non-centrosymmetric stacking facilitates stronger nonlinear interactions, thereby enhancing entangled photon generation efficiency. Gao and co-workers[104] have experimentally demonstrated the tunability of EPPs emission in 2D r-BN, a vdW insulating material. The quantum state of the generated entangled photon pairs actively controlled by manipulating the polarization orientation of the incident pump field. In this context, hexagonal 2D r-BN is investigated under a polarized optical pump field. The resulting 2nd order nonlinear polarization response in the material is described by the relation

$$P_{\alpha}^{2\omega} = \varepsilon_0 \chi_{\alpha\beta\gamma}^{(2)} E_{\beta}^{\omega} E_{\gamma}^{\omega} \tag{10}$$

where $\varepsilon_0$ is the free space permittivity, $\chi_{\alpha\beta\gamma}^{(2)}$ is the second order nonlinear susceptibility (SONS) and $\alpha, \beta, \gamma$ is $x$ or $y$ is the coordinates, $E_{\beta}^{\omega}$ and $E_{\gamma}^{\omega}$ are the components of the electric field of the pump at frequency of $\omega$. In a monolayer h-BN the nonvanishing SONS is $\chi_{yyy}^{(2)} = -\chi_{yxx}^{(2)} = -\chi_{xxy}^{(2)} = -\chi_{xyx}^{(2)} = |\chi^{(2)}|$, where x and y corresponding to the armchair and gig-zag direction of the crystal.

The six-fold symmetry observed in the SPDC signal response exhibits periodic maxima and minima as a function of the pump polarization angle. Under co-polarized detection conditions, the SPDC signal reaches a maximum when the pump polarization is aligned along the AC direction of the r-BN crystal, and a minimum when aligned with the ZZ direction (**Figure 3m**). This behavior is the dominance of the nonzero tensor component $\chi_{yyy}^{(2)}$ in second-order nonlinear susceptibility, as indicated in **equation 10**. The six-fold symmetry remains preserved under cross-polarized pump excitation, except for a phase shift of 30°, which is a symmetry rotation due to tensorial selection rules. Notably, when the pump field is aligned along the zigzag direction under cross-polarized conditions, the photon pair generation is enhanced, corresponding to the contribution from the non-zero component $\chi^{(2)}_{xyx}$. In a complementary configuration, the analyzer is rotated while the pump polarization is held fixed (**Figure 3o-q**). The resulting two-lobed symmetrical patterns are interpreted based on the quantum state probability amplitude

$$\Phi = \frac{\mathrm{Cos}(\varphi_p)}{\sqrt{2}}(|HV\rangle + |VH\rangle) + \frac{\mathrm{Sin}(\varphi_p)}{\sqrt{2}}(|HH\rangle - |VV\rangle) \quad (11)$$

where $\varphi_p$ denotes the pump polarization angle. When the analyzer is set at an angle θ, the resulting quantum states exhibit characteristic angular dependencies. The coincidence count rate follows the analyzer angle, exhibiting modulation proportional to $\mathrm{Sin}^2(\varphi_p+2\theta)$, $\mathrm{Sin}^2\varphi_p$, or zero, depending on the specific analyzer pump angle configuration. These angular dependencies reflect the underlying polarization entanglement and the anisotropic nonlinear optical response of the r-BN crystal. The rotational dependence of the pump polarization allows for tunable modulation of the EPPs characteristics, making r-BN a promising platform for polarization engineered quantum light sources.

On another report the entangled photon pairs that are generated in r-BN for the on chip integrated quantum applications is reported. The polarization analysis of generated photon pairs is explained with crystal symmetry. The system is defined with special polarization relationship between the pump, signal and idler light is

$$|R\rangle_p \xrightarrow{\mathrm{SPDC}} |L\rangle_s|L\rangle_i \quad (12a)$$

$$|L\rangle_p \xrightarrow{\mathrm{SPDC}} |R\rangle_s|R\rangle_i \quad (12b)$$

$$|H\rangle_p \xrightarrow{\mathrm{SPDC}} \frac{1}{\sqrt{2}}(|H\rangle_s|H\rangle_i - |V\rangle_s|V\rangle_i) \quad (13a)$$

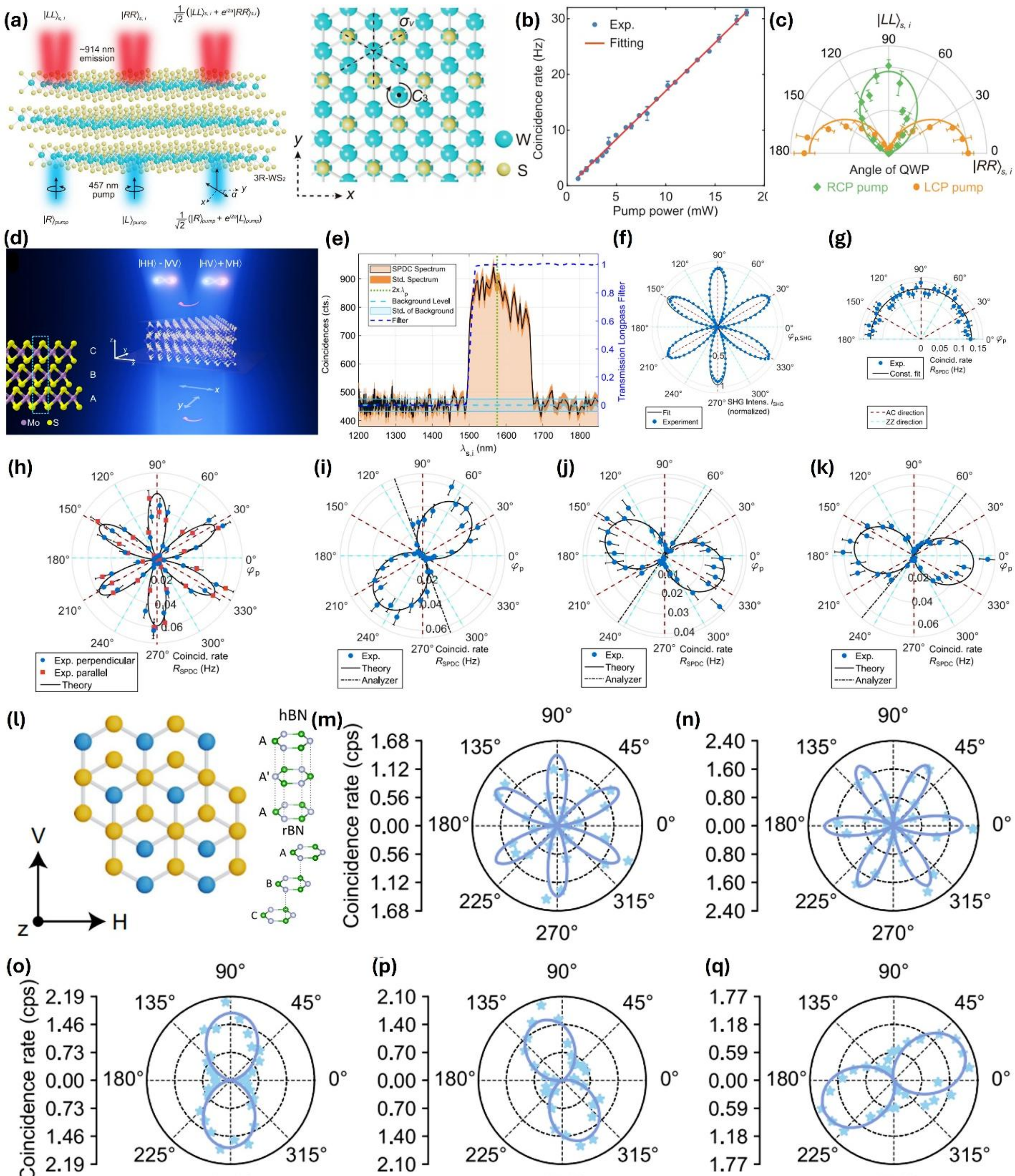

**Figure 3| Quantum entangled photon pairs from a van der Waal semiconductors and insulators. a,** *Left:* Conceptual illustration of the generation of polarization-entangled photon pairs in a thin crystal of 3R-$WS_2$. *Right:* Schematic representation of 3R-$WS_2$. The symbols $C_3$ and $\sigma_v$

represent the 3-fold RS axis and the mirror symmetry plane of the crystal, respectively. **b,** Coincidence rate of generated EPPs as a function of pump power. **c,** CP dependent photon pair generation. Left- and right-circularly polarized pump dependent photon pairs generation and the coincidence rate as a function of the analyzer angle. Adopted from Springer Nature.[37] Copyright, 2024, Creative Commons Attribution 4.0 International License. **d,** The entangled photon pair generation in a 3R-$MoS_2$ multilayer. Polarization-entangled *signal* and *idler* photons are generated via SPDC. Maximally entangled Bell states are obtained depending on the orientation of the pump polarization and ABC stacking scheme of the crystal. **e,** Measured SPDC photon spectrum from 3R-$MoS_2$. The SPDC wavelength is $\lambda_{deg} = 2 \times \lambda_p = 1576$ nm**. f,** P-SHG of the 3R-$MoS_2$ crystal, theoretically fitted with $I_{SHG} \propto Sin^2(3\varphi_{p,SHG})$, measured along the armchair (AC) and zigzag (ZZ) directions of the crystal. **g,** Unpolarized detection rate of EPPs from SPDC with rotating pump polarization. **h,** Polarization-dependent detection rate of EPPs from SPDC with rotating pump polarization. The SPDC dependence is $R_{SPDC} \propto Sin^2(3\varphi_p)$. **i-k,** Photon pair detection rate with fixed analyzers under rotating pump polarization ($\varphi_p$). The analyzer angles are fixed at 110°, 55°, and 50° for configurations ***i***, ***j***, and ***k***, respectively. The SPDC dependence is as follows $R_{SPDC} \propto Sin^2(3\varphi_{pol} + \varphi_p)$. Adopted from Springer Nature.[64] Copyright, 2024, Creative Commons Attribution 4.0 International License. **l,** *Left:* Schematic of the r-BN crystal structure. *Right:* Comparison of the crystal structures and stacking sequences of h-BN and r-BN. The armchair (V) and zig-zag (H) directions of the crystal are indicated. **m,** Coincidence count rate as a function of incident pump polarization in the co-polarization configuration, where the polarization of both collection arms is aligned parallel to the incident pump polarization. **n,** Coincidence count rate in the cross-polarization configuration, where one collection arm is aligned parallel and the other perpendicular to the incident pump polarization. **o-q,** Pump polarization dependent coincidence rates with fixed analyzer angles in the collection arms. Adapted from.[104] Copyright 2025, Springer Nature.

where R and L are represented the horizontal (H) and vertical (V) polarization basis. The above equation of the polarization evolution process of SPDC becomes

$$|V\rangle_p \xrightarrow{\text{SPDC}} \frac{-1}{\sqrt{2}}(|H\rangle_s|V\rangle_i - |V\rangle_s|H\rangle_i) \tag{13b}$$

When the pump is arbitrarily linear polarized then the SPDC is projected as a

$$\frac{-Cos\varphi_p}{\sqrt{2}}(|H\rangle_s|V\rangle_i - |V\rangle_s|H\rangle_i) + \frac{-Sin\varphi_p}{\sqrt{2}}(|H\rangle_s|H\rangle_i - |V\rangle_s|V\rangle_i) \tag{14}$$

where $\varphi_p$ is the angle between the pump and the V polarization direction. A constant SPDC coincidence rate is observed as varied pump polarization, even in the absence of a polarizer at the detector (**Figure 4a**). In contrast, distinct six-fold symmetric patterns emerge when both the pump and analyzer polarizations are kept parallel and rotated simultaneously (**Figure 4b**). The observed maxima and minima in the SPDC coincidence rate correspond to the armchair and zig-zag directions of the crystal plane, respectively, indicating that the crystal orientation plays a crucial role in determining the coincidence generation rate.

When the analyzer is aligned along the zig-zag direction (0°) and the pump polarization is set along the armchair direction, the coincidence rate follows a $Sin^2\varphi_p$, dependence (**Figure 4c**). Furthermore, the polarization dependent coincidence rate shown in **Figures 4c-f** exhibits a dependence proportional to $Sin^2(\varphi_p + 2\varphi_{pol})$. By fixing the pump polarization and rotating the analyzer angle, the coincidence rate varies as $Sin^2(2\varphi_{pol} + \varphi_p)$, as demonstrated in **Figures 4g and h**.

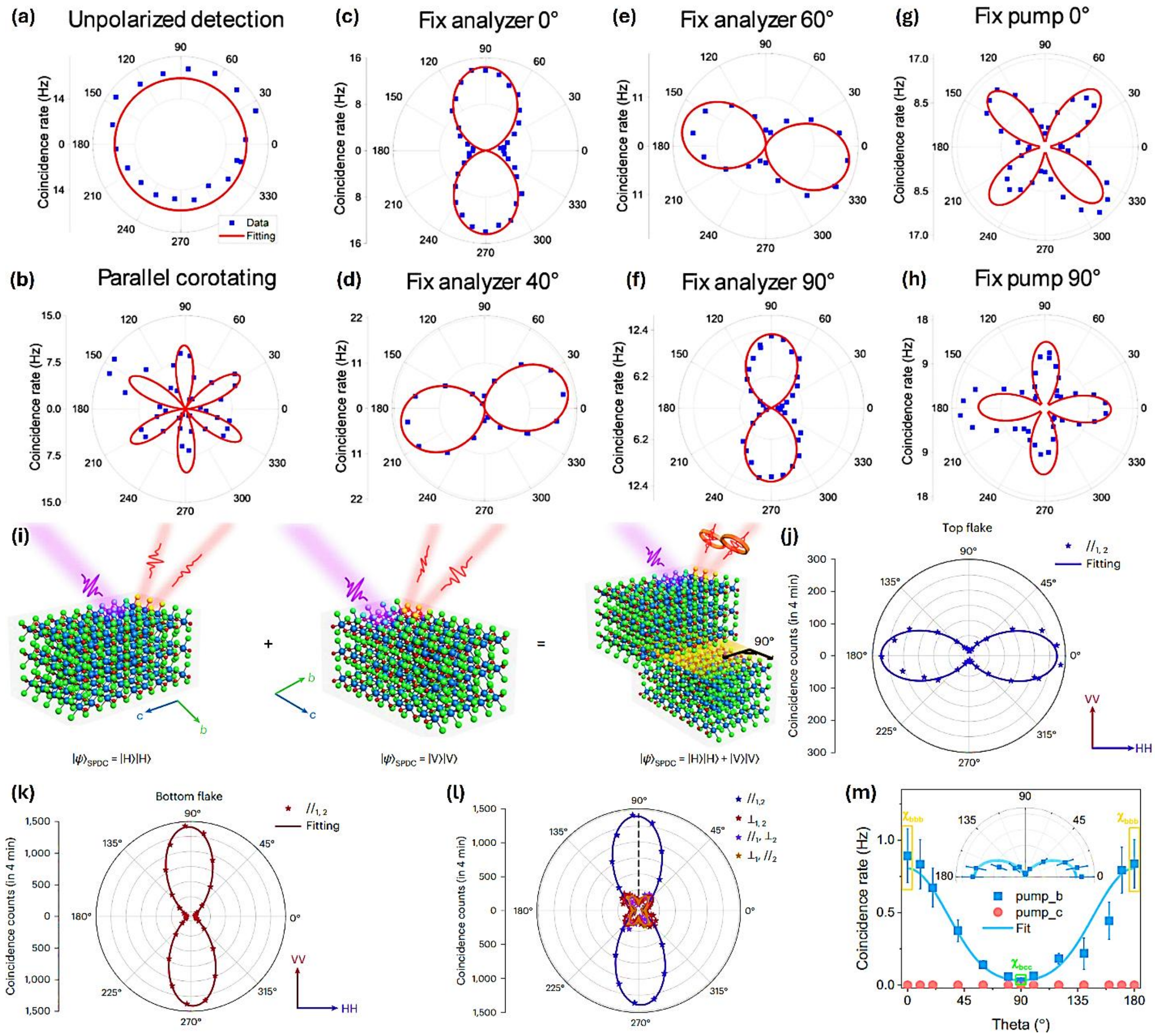

**Figure 4| Photon pair generation and polarization analysis in r-BN, $NbOCl_2$ heterostructures. a,** Coincidence count rate as a function of pump polarization angle ($\varphi_p$), measured without any polarization filters at the detectors. **b,** Coincidence rate when both the pump and detector polarizations are rotated simultaneously ($\varphi_p = \varphi_{pol}$). The dependence follows $R_{SPDC}\ \alpha\ Sin^2(3\varphi_p)$. **c-f,** Coincidence rates as a function of $\varphi_p$ with fixed analyzer (detector)

polarization angles $\varphi_{pol}$ set at 0°, 40°, 60°, and 90°, respectively. The dependence follows $R_{SPDC}\ \alpha\ Sin^2(2\varphi_{pol} + \varphi_p)$. **g, h,** Coincidence rates with fixed $\varphi_p$ at 0° and 90°, respectively, as the analyzer polarization angle $\varphi_{pol}$ is varied. The dependence behavior is described by the relation $R_{SPDC}\ \alpha\ Sin^2(2\varphi_{pol} + \varphi_p)$. Reproduced from Liang *et al.*,[17] Copyright 2025, AAAS. **i**, Conceptual illustration of entangled photon pair generation using a two-crystal geometry in $NbOCl_2$, a 2D vdW material. **j, k**, Pump polarization dependent coincidence rates measured from the top and bottom layers of the crystal, respectively. **l,** All polarization mode pump polarization dependent coincidence. The '//' and '⊥' sign denotes the co-polarized and cross-polarized to the pump beam in the corresponding collection arms 1 and 2, respectively. Reproduced with permission from.[105] Copyright 2024, Springer Nature. **m**, Polarization dependent photon pair coincidence rate in $NbOCl_2$ with error bars. The inset shows the corresponding polar coordinate plot. Adapted from.[23] Copyright 2024, Springer Nature.

Beyond BN, the heterostructure of two-crystal geometry approach is employed to generate entangled photon pairs.[105] In this method, two perpendicular vdW crystals of 2D niobium oxide dichloride ($NbOCl_2$) are stacked (**Figure 5i**). In such a stacked crystal, type-I phase matching is achieved, with the crystals aligned so that their optical axes are perpendicular to each other. Entangled photon pairs are generated via SPDC occurring in either of the two crystals. The indistinguishability of the down conversion process between the two orientations leads to the creation of an entangled photon pair is

$$|\emptyset^+\rangle = |H\rangle|H\rangle + |V\rangle|V\rangle \tag{15}$$

The states $|H\rangle|H\rangle$ and $|V\rangle|V\rangle$ correspond to photon pairs generated along the optical axes of the first and second crystals, producing horizontally and vertically polarized photon pairs, respectively. In this system, coincidence count depends on the pump polarization and exhibits strong directional dependence. Specifically, the peak coincidence counts for one flake occurs when the pump polarization is perpendicular to that of the other flake (**Figure 5j,k**). Furthermore, a remarkably high nonlinear coefficient has been achieved, along with polarization dependent on enhancements in the coincidence rates. Nonetheless, the single-flake $NbOCl_2$ exhibited a pump-polarization-dependent coincidence rate (**Figure 5l)**. Orthogonally stacked (OS) van der Waals (vdW) crystals have been used to explore polarization-dependent entanglement.[23] The OS crystal structure exhibits a polarization-dependent coincidence rate (PDCR) (**Figure 5m**), which is consistent with the crystal's SHG response.

## 4. Quantum state tomography and applications

Quantum state tomography (QST), also known simply as quantum tomography (QT), is a fundamental technique in quantum information science for determining the complete description of an unknown quantum state. It involves subjecting a large ensemble of identically prepared quantum systems to a series of measurements in different, non-commuting bases. Through these measurements, the quantum state can be reconstructed using the principles of quantum mechanics. QST plays a vital role in verifying the preparation of target quantum states, benchmarking quantum sources, and validating protocols for quantum communication, computation, and metrology, making it an essential tool in various areas of quantum technology. The QST is used to measure the generated polarization quantum state and to demonstrate entanglement between the signal and idler photons in 2D quantum material systems.[23,105–109]

The first demonstration of quantum states in 2D vdW materials was realized in 3R-$WS_2$.[37] A complete quantum state tomography (QST) was performed, enabling the reconstruction of the density matrix of two-photon states. In this experiment, the signal and idler photons were separated using a polarization-independent beam splitter. Notably, entangled photon pairs were generated using LCP and RCP pump beams. The polarization density matrix $\hat{\rho}$ of the quantum state was then fully determined by performing projections onto 16 distinct basis states (**Figure 5a, b**).[110] On the other hand, two maximally entangled Bell states are constructed. The polarization-entangled state is given by $\frac{1}{\sqrt{2}}(|LL\rangle_{s,i} + e^{i2\alpha}|RR\rangle_{s,i})$, which is achieved using linearly polarized pump light of the form $\frac{1}{\sqrt{2}}(|R\rangle_{Pump} + e^{i2\alpha}|L\rangle_{Pump})$ at an angle $\alpha$. When $\alpha = 0$, the quantum state becomes $\frac{1}{\sqrt{2}}(|LL\rangle_{s,i} + |RR\rangle_{s,i})$ and the corresponding Bell state is $\frac{1}{\sqrt{2}}(|HH\rangle_{s,i} + |VV\rangle_{s,i})$. When $\alpha = \frac{\pi}{2}$, the photon-pair state is $\frac{1}{\sqrt{2}}(|LL\rangle_{s,i} + |RR\rangle_{s,i})$ and the Bell state becomes $\frac{1}{\sqrt{2}}(|HV\rangle_{s,i} + |VH\rangle_{s,i})$. These two states in 3R-$WS_2$ are prepared using linearly polarized light, and the corresponding density matrices are shown in **Figures 5c, d**. The QST results for 3R-$WS_2$ closely match the Bell states, exhibiting fidelities of 0.93 and 0.96, and concurrences of 0.97 and 0.98 for the two Bell states, respectively.

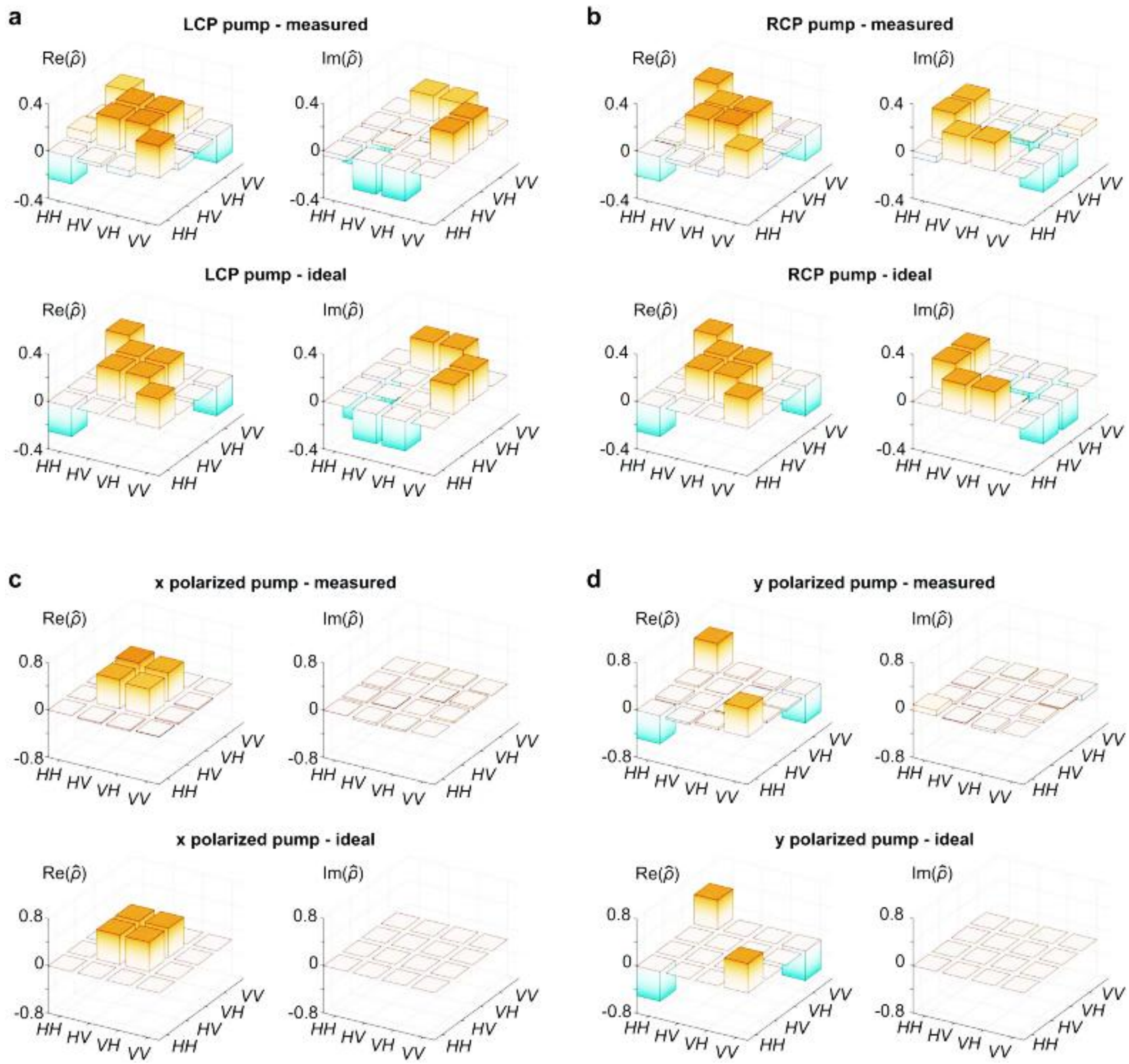

**Figure 5| Quantum state tomography of P-EPPs in $WS_2$.** The quantum state is fully characterized by its density matrix. **a**, the $|RR\rangle_{s,i}$ state pumped with LCP light; **b**, the $|LL\rangle_{s,i}$ state pumped with RCP light; **c**, the Bell state generated with LP pump light along the zigzag direction (*x*-axis); and **d**, the Bell state generated with LP pump light along the armchair direction (*y*-axis). Adopted from Springer Nature.[37]

In some cases, an established maximum likelihood method is employed to determine a physically valid density matrix $\hat{\rho}$. The real and imaginary components of the density matrix are extracted from various vdW materials, as shown in **Figure 6.** For instance, Kallioniemi et al.[105] reconstructed a density matrix that closely resembles the expected quantum state (**Figure 6a**). The results demonstrate a fidelity of 86±0.7%, indicating a high level of quantum state engineering. Concurrently, concurrence C is a measure of photon state entanglement, with values ranging from 0 to 1. Specifically, C=0 corresponds to a completely separable (non-entangled) state, whereas C=1 indicates a maximally entangled state.[111,112] For the $NbOCl_2$ system, the estimated concurrence is 0.87±0.010. In another report, the polarization state of photon pairs generated in a single layer (1L) $NbOCl_2$ crystal was investigated.[23] Notably, the polarization of the generated photon pairs is independent of the pump polarization and followed an HH state, where *H* denotes polarization along the *b*-axis of the crystal (**Figure 6b**). These HH-state photon pairs exhibited a high fidelity of 0.95 ± 0.02. Importantly, the generation of photon pairs in the 1L crystal is constrained by the crystal's symmetry, which is inherently polarization independent. As a result, the system does not exhibit polarization entanglement. This observation directly correlates with the structure of the crystal's nonlinear susceptibility tensor. The nonlinear optical properties of a material are fundamentally governed by its crystal symmetry,[19,113–115] which in turn influences the SPDC process and the quantum states of the generated EPPs. Experimental quantum state tomography of a vdW 3R-$MoS_2$ crystal has been reported, with the polarization resolved density matrices for *x*- and *y*- polarized excitation shown in **Figure 6c** and **6d**, respectively. Under *y*-polarized excitation, the generated quantum state exhibits a high fidelity of 0.96 with the maximally entangled Bell state $|\Psi\rangle = \frac{1}{\sqrt{2}}(|HH\rangle - |VV\rangle)$. For *x*-polarized excitation, the fidelity is slightly lower, at 0.84, corresponding to the state $|\Psi\rangle = \frac{1}{\sqrt{2}}(|HH\rangle + |VV\rangle)$. Additionally, the

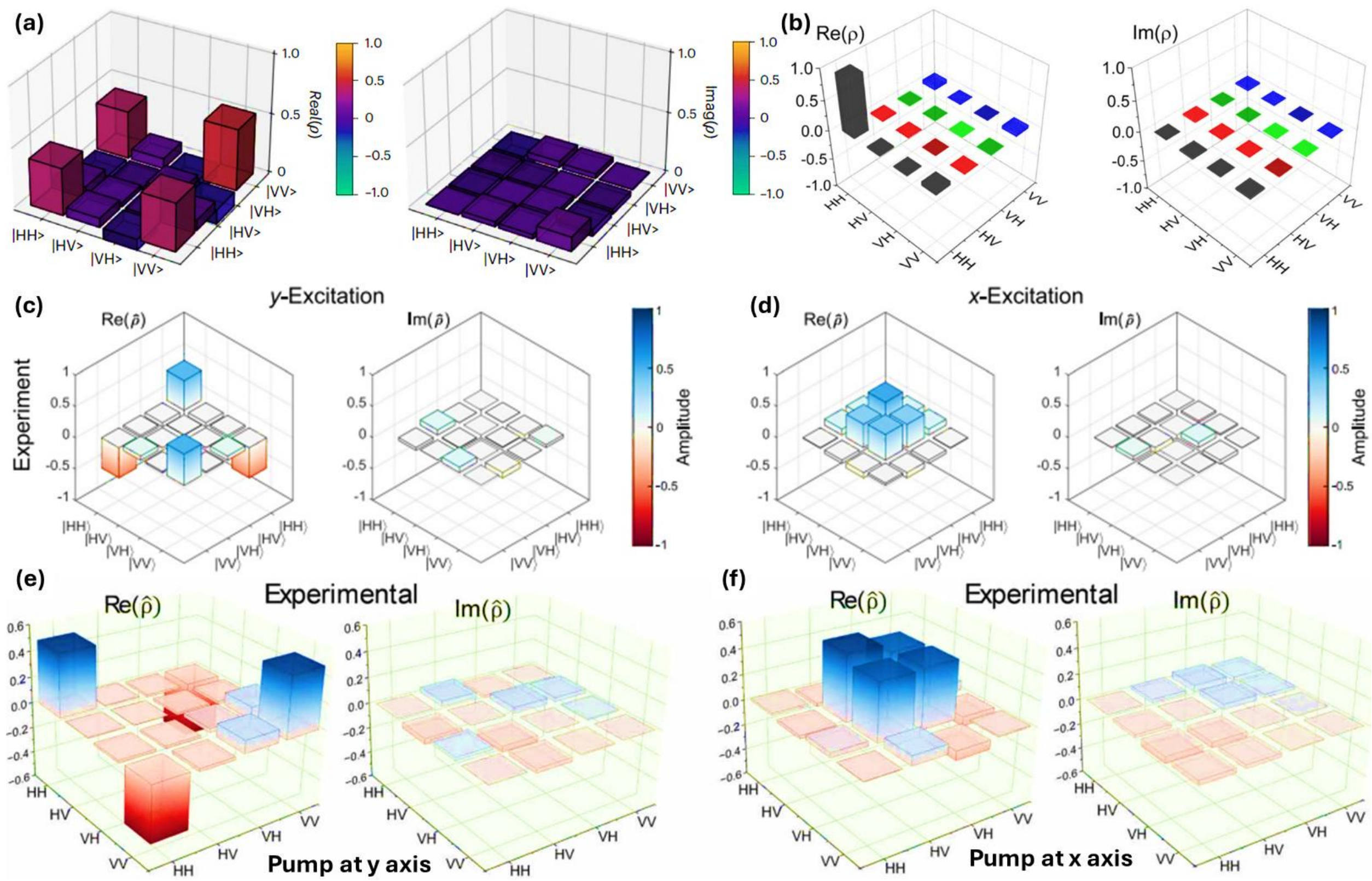

**Figure 6| Entangled quantum states. a,** Quantum state tomography of the entangled photon pairs. *Left:* real part; *right:* imaginary part of the reconstructed density matrix. Reproduced with permission from.[105] Copyright 2024, Springer Nature. **b**, Polarization dependent quantum state tomography in a single flake. Density matrices correspond to a fixed pump polarization of 45°. Adapted from Guo et al.[23] Copyright 2024, Springer Nature. **c**, Polarization density matrices for *y*-polarized pump; **d**, for *x*-polarized pump. The real and imaginary parts are denoted by Re(ρ̂) and Im(ρ̂), respectively. The results obtained under *x*- and *y*- polarized excitation of 3R-$MoS_2$. Adapted from Weissflog et al.[64] Copyright 2024, Springer Nature. **e**, Density matrix of the generated photon quantum state under a pump polarization angle of 90°, corresponding to the entangled state $|\Psi\rangle = \frac{1}{\sqrt{2}}(|HH\rangle - |VV\rangle)$, **f**, under a pump polarization angle of 0°, corresponding to $|\Psi\rangle =$

$\frac{1}{\sqrt{2}}(|HH\rangle + |VV\rangle)$. Reproduced from Liang et al.,[17] Copyright License 4.0 (CC BY- NC) 2025, AAAS.

measured concurrences $C$ are 0.973 ± 0.002 for $x$-polarization and 0.82 ± 0.02 for $y$-polarization.[23] Such high concurrence values confirm the presence of strong quantum correlations between the photon pairs. Particularly, a concurrence value approaching unity signifies a high degree of entanglement between two qubits.

The polarization quantum state of *C3* symmetric r-BN has been explored to produce entanglement between signal and idler photons.[17] Quantum state tomography performed with a pump polarization angle of 90° reveals an entangled state described by $|\Psi\rangle = \frac{1}{\sqrt{2}}(|HH\rangle - |VV\rangle)$ with the corresponding density matrix shown in **Figure 6e**. In contrast, when the pump polarization is set to 0°, the generated state is $|\Psi\rangle = \frac{1}{\sqrt{2}}(|HH\rangle + |VV\rangle)$ as depicted in **Figure 6f**. The obtained fidelities for these states are 0.97 ± 0.013 and 0.933 ± 0.002 for the 90° and 0° pump polarizations, respectively. These high-fidelity values promise that the prepared quantum states in r-BN are in excellent agreement with the ideal Bell states. Additionally, the strong entanglement quality is confirmed by the high concurrence values of the photon pairs. A summary of the most recent quantum state tomography results for entangled photon pairs in various vdW materials is presented in **Table 2**. In addition, the crystal symmetry of vdW materials plays a crucial role in influencing the EPP generation rate. **Figure 7a** illustrates the EPP generation rates for various 2D vdW materials. Notably, r-BN exhibits an exceptionally high EPP generation rate compared to other layered materials. The combination of outstanding quantum tomographic performance and high EPP generation rates underscores the strong potential of these materials for next generation, practical on-chip quantum technologies.

**Table 2.** *Entangled photon pairs generated signals in vdW materials and their physical parameters. CAR is Coincidence to accidental ratio.*

| Materials | Polarization Tomography Type | Photon pair generation rate | Fidelity | CAR | Reference |
|---|---|---|---|---|---|
| r-BN | 2-photon | 120 Hz | 0.93 | >200 | [17] |
| $MoS_2$ | 2-photon | 1563 | 0.96 | 0.973 | [64] |
| r-BN | - | 8667 Hz/(mW·mm) | 94% | 2700 | [104] |
| $NbOCl_2$ | 2-photon | 31 kHz (mW·mm) | 86% | 1500 | [105] |
| SiC | - | - | 0.97 | - | [116] |

## 5. On-chip quantum device applications

On-chip quantum technologies are leading the advancement of industrial quantum applications.[3,33,34,117,118] In particular, integrated quantum photonics offers a compact, reliable, reprogrammable, and scalable platform for both the exploration of fundamental quantum physics and the development of advanced quantum technologies. While bulk nonlinear crystals have been explored for on-chip device applications especially nonlinear photonic devices, vdW materials are now being considered as promising candidates for on-chip nonlinear photonic devices tailored for quantum applications. One key challenge lies in integrating SHG into existing silicon photonics platforms, primarily due to the lack of CMOS compatibility with conventional nonlinear materials.

Additionally, the fixed and nontunable response of current nonlinear optical (NLO) materials significantly limits their practical implementation.

Most recently, an on-chip light source embedding 2D TMDs for integrated photonics has been realized.[34] In particular, a layered 2D vdW $MoSe_2$ flake was embedded onto a plasmonic nano waveguide using a viscoelastic transfer technique (**Figure 7b**). In this device, the emission from the layered $MoSe_2$ can be directly coupled, on-chip, into plasmonic slot waveguides (**Figure 7c**). The reference spectrum of $MoSe_2$ from a sample without any waveguide or nanoantenna, compared with the waveguide output, shows a significant, red-shifted feature. Corresponding images of the waveguide output were also captured. Moreover, polarization-dependent measurements confirm the bright-spot features at the output coupling antennas (**Figures 7d**), demonstrating efficient optical coupling between the antenna gap plasmon resonance and the waveguide mode. On the contrary, Tang et al.[33] introduced an on-chip device platform for 2D materials with in situ adjustable interfacial properties. The device was realized using a microelectromechanical system (MEMS), in which an hBN crystal was interfaced with a silicon photonic platform (**Figure 7e**), enabling compact and cost-effective integration. This was achieved by precisely controlling the hBN layers through twisting and pressurizing actions. SHG measurements revealed that the vertical control voltage $V_z$ could tune the vertical gap h between the layers (**Figure 7f**). The tunability of SHG in this system was more than an order of magnitude higher than that observed in a Fabry-Perot cavity composed of two hBN layers. In this platform, quantum entanglement was realized through SPDC, where one photon at 2ω is converted into two photons at ω, inherently generating entangled photon pairs with high purity. The system can generate entangled photon pairs with adjustable entanglement properties, enabling a tunable

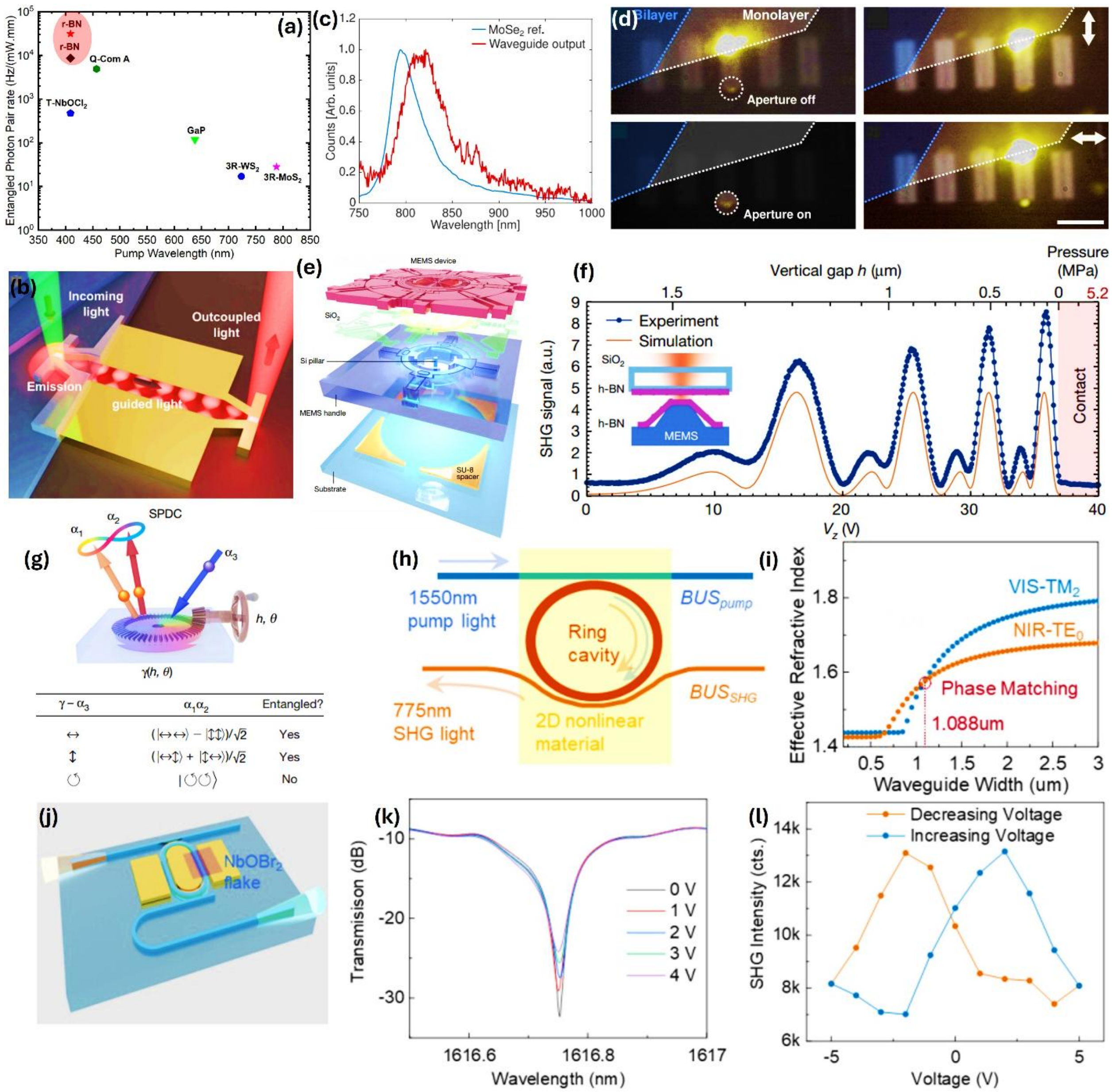

**Figure 7| a,** A comparison of entangled photon pair generation rates in various vdW materials.[37,64,104,105,119,120] Among them, r-BN exhibited an excellent entangled photon pair generation rate. **b,** Schematic of the experimental setup used to stimulate photoluminescence in $MoSe_2$ attached to a plasmonic waveguide device, with the output coupled to free space via a terminal dipole antenna. **c***,* PL spectra of the reference $MoSe_2$ sample and the output of the waveguide. **d***,* top and bottom left: output of the waveguide showing only the measured outcoupled

emission; top and bottom right: waveguide emission recorded with a polarization filter at cross and parallel polarizations. Adapted with permission.[34] Copyright 2025, American Chemical Society. **e**, Schematic of the main components of a fabricated MEMS-based Generic Actuation platform for 2D materials (MEGA2D) device. **f**, SHG power of the h-BN–integrated MEGA2D device as a function of $V_z$. **g**, SPDC from the fabricated MEGA2D devices, demonstrating the potential for entangled-state generation. Adapted with permission.[33] **h,** Schematic of on-chip SHG device. **i,** Refractive index of the SiN waveguide and the phase matching curve, **j,** Proposed tunable SHG device, **k,** Applied voltage dependent device transmission spectra, **l,** Cyclic applied voltage dependent SHG intensity of the device. Adopted with permission from authors.[117] Copyright SPIE, 2025.

quantum source (**Figure 2g**). In particular, if $\alpha_1 + \alpha_2 = \gamma - \alpha_3 = 0$ with a Bell State is $\frac{(|\leftrightarrow\leftrightarrow\rangle - |\updownarrow\updownarrow\rangle)}{\sqrt{2}}$ and if $\alpha_1 + \alpha_2 = \frac{\pi}{2}$ another Bell State is $\frac{(|\leftrightarrow\updownarrow\rangle - |\updownarrow\leftrightarrow\rangle)}{\sqrt{2}}$. If $\alpha_1$ and $\alpha_2$ are forced to zero then the generated photon has no entanglement ($| \circlearrowleft\circlearrowleft\rangle$, product state)."

Nonetheless, EPP based on vdW materials have yet to be explored for on-chip quantum device integration. However, the advances in on-chip nonlinear optics have been significantly propelled by the integration of 2D materials exhibiting strong second-order nonlinearities. A notable development is the use of niobium oxide dibromide ($NbOBr_2$) in silicon photonics platforms, as demonstrated by Gong et al.,[117] where the material's intrinsic centrosymmetry breaking structure and in-plane ferroelectricity enable efficient and tunable SHG (**Figure 7h-l**). Unlike traditional TMDs and layered materials whose SHG responses are layer dependent, $NbOBr_2$ offers robust, layer number independent SHG due to its anisotropic lattice distortions. When integrated into a SiN microring resonator, $NbOBr_2$ facilitates high efficiency SHG ($\eta_{SHG}$) under low power continuous wave excitation (**Figure 7l**), aided by cavity enhanced field confinement. More critically, the authors exploit the ferroelectric domain switching of $NbOBr_2$ to achieve nonvolatile electrical tuning of the SHG response, a novel mechanism that introduces hysteresis and memory effects into integrated nonlinear photonic devices (**Figure 7k, l**). This multifunctional approach merges frequency conversion with programmable optical modulation, presenting a promising route toward compact, reconfigurable photonic circuits and next-generation optics-based quantum computing technologies.

## 6. Summary with Challenges and Outlook

The review article outlines the fundamental properties of EPP generation, their characterization, and the purity of entangled photons in 2D vdW materials. It summarizes the nonlinear quantum optical properties of various 2D vdW materials and highlights the influence of crystallographic structures, especially crystal symmetry, on these properties. The NLO behavior of 2D vdW materials, which underpins the SPDC process for generating EPPs, is explored in detail. Different crystallographic symmetries in layered materials reveal a wide range of $2^{nd}$ order nonlinear responses. For example, r-BN and 3R-$MoS_2$ exhibit rich nonlinear responses, enabling the generation of quantum EPPs via SPDC. Strong quantum correlations between EPPs have been confirmed through quantum state tomography. A high degree of entanglement between qubits is evident in the coincidence measurements performed on vdW materials. However, the realization of qubits with long coherence times and reduced noise remains largely unexplored. Nevertheless, experimental research on entangled photon generation in atomically thin layers and their on-chip integration for future quantum applications is still in its very early stages.

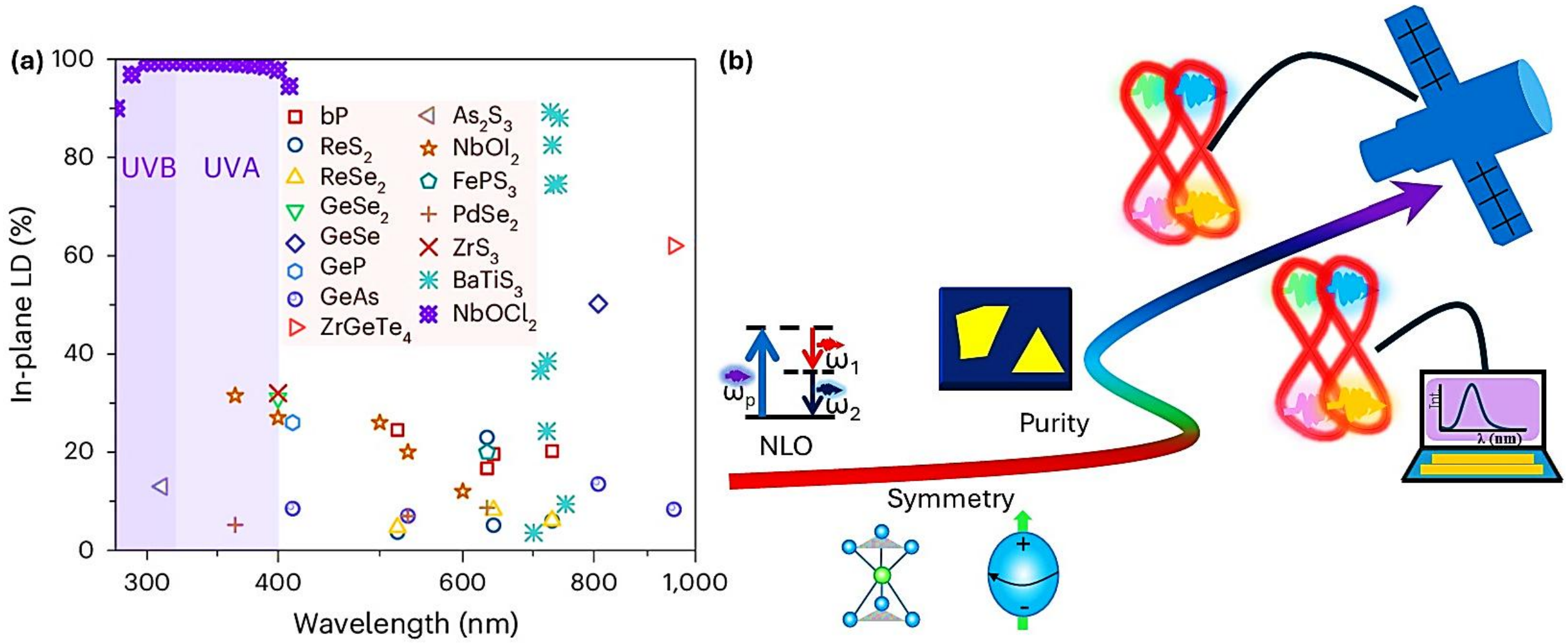

**Figure 8|** ***a,*** In-plane linear dichroism in recently developed vdW materials. Adopted with permission from.[121] Copyright 2024, Springer Nature. and ***b***, Schematic illustration of EPPs generations in vdW materials for future quantum photonic and quantum opto- electronic applications.

The generation of EPPs in nonlinear materials is reshaping the future of present quantum technology. Recent advancements are no longer confined to conventional, widely used nonlinear bulk crystals but are progressively shifting towards quantum confined vdW materials.[122] The boosting of classical and quantum nonlinear processes in layered vdW materials have emerged at the forefront of current research.[25,112,123,124] Starting from layered graphene, the field now includes transition metal mono-, di-, and tri-chalcogenides, perovskites, metal-organic frameworks (MOFs), and singlecrystal molecular systems all of which exhibit remarkable nonlinear and quantum optical properties.[98,125–133] Several of these materials have demonstrated exceptionally high second order ($\chi^{(2)}$) and higher order ($\chi^{(n)}$, n ≥ 3) nonlinear coefficients.[19,43,134–136] TMDs, in particular, have shown promising capabilities in producing EPPs with excellent associated parameters. Moreover, in-plane anisotropic layered materials exhibit the lowest crystal symmetry among known layered structures (**Figure 8**). Materials such as black phosphorene and monochalcogenides including SnS, SnSe, GeS, and GeSe possess rich anisotropic nonlinear coefficients.[75,129,137–139] Additionally, two-dimensional phase change materials may hold the potential to revolutionize entangled photon pair generation due to their extraordinary second order nonlinear susceptibility.[140–142] Beyond this, high dimensional quantum frequency combs[143] in 2D vdW materials remain largely unexplored, potentially opening new avenues for quantum optics generation and device applications.

From a materials perspective, the application of 2D vdW materials in quantum entanglement and thus in quantum optics for on-chip quantum applications requires the growth of wafer scale, high crystal quality materials. Achieving high quality crystal growth presents challenges, as crystallinity and defects in thin layer 2D materials are critical issues. Although, the growth of specific crystal phases of atomically thin materials on appropriate substrates is a key focus area demanding substantial effort. In this context, more scalable and reliable synthesis methods capable of fabricating vdW materials and heterostructures with nanometric precision are urgently needed for on-chip quantum device applications. Moreover, the discovery of EPPs in novel materials, such as magnetic 2D materials including anisotropic ones, remains largely unexplored.[144–146] In addition, vdW heterostructures (HS), including those with mixed dimensionalities and crystal symmetries, can reduce PL induced background noise and hence can enhance the second order nonlinear susceptibility $\chi^{(2)}$, thereby improving SPDC and the generation of entangled photon pairs. Besides this, the low CAR value in vdW materials limits the scalability and integration of photonic quantum applications. This is primarily the emission from the material over a broad wavelength range. The twisted phase-matching (TwPM) model introduces a nonlinear geometric phase from the twisted bilayers, compensating for the phase difference in SHG. This enables manipulation of the electronic band structure and the corresponding optical emission characteristics. This TwPM approach may directly enhance the CAR value of entangled photon pairs.

Another critical challenge lies in the use of vdW materials in in-plane incidence geometries for on-chip device applications. So far, only normal-incidence geometries have been explored for entangled photon-pair sources, where the corresponding quantum optical properties have been exploring recently. However, the in-plane quantum optical properties-including SPDC quasi-phase matching and, most importantly, the quantum coherence length-remain largely unexplored. Investigating these properties could significantly advance the field and help overcome many practical challenges in integrating such ultrathin layered vdW sources into on-chip quantum device applications.

Nonetheless, a central objective in quantum information science is the generation and manipulation of more complex entangled states. In particular, multiphoton entangled quantum states are currently regarded as core resources for advancing the field. In this context, the development of large-scale entangled quantum states in 2D vdW materials still lags behind. Another pressing challenge is integrating high purity, high efficiency, and high brightness into a single photonic quantum source, alongside accurately estimating and minimizing various losses. Furthermore, promising directions for future research include the realization of integrated SPDC sources and high-dimensional multiphoton sources based on vdW materials.

Finally, while challenges remain, the outlook for vdW materials in entangled photon generation is overwhelmingly positive. Their unique crystal symmetries, structures, and nonlinear properties, combined with cutting edge innovations, position them as transformative building blocks for the next generation of quantum technologies. Nonetheless, the generation of EPPs in ultrathin layered vdW materials is expected to enable advancements in quantum communication, quantum computing, quantum sensing for biological sciences, the semiconductor industry, and on-chip consumer applications.

**Acknowledgements**. The author is thankful for the kind permission from the corresponding publishers or authors to reproduce the figures included in this article.

**Conflict of Interest.** The authors declare no conflict of interest.